\documentclass[aps, prd, twocolumn, superscriptaddress, showkeys, byrevtex groupedaddress]{revtex4-1}
\usepackage{scalerel}
\usepackage{dcolumn}   
\usepackage{bm}        
\usepackage{amssymb, amsmath,array}   
\usepackage{graphicx,float}
\usepackage{subfig, caption}
\usepackage{url}
\usepackage{natbib}
\usepackage{mathtools}
\usepackage{txfonts}
\usepackage[normalem]{ulem}
\usepackage{booktabs}
\usepackage{epsfig}
\usepackage{hyperref}
\hypersetup{
	colorlinks=true,
	citecolor=blue,
	linkcolor=blue,
	urlcolor=blue}
\usepackage{tikz,xcolor}
\definecolor{lime}{HTML}{A6CE39}
\DeclareRobustCommand{\orcidicon}{%
    \begin{tikzpicture}
    \draw[lime, fill=lime] (0,0) 
    circle [radius=0.16] 
    node[white] {{\fontfamily{qag}\selectfont \tiny ID}};
    \draw[white, fill=white] (-0.0625,0.095) 
    circle [radius=0.007];
    \end{tikzpicture}
    \hspace{-2mm}
}
\foreach \x in {A, ..., Z}{%
    \expandafter\xdef\csname orcid\x\endcsname{\noexpand\href{https://orcid.org/\csname orcidauthor\x\endcsname}{\noexpand\orcidicon}}
}

\begin{document}
	\title{Baryon-Dark matter interaction in presence of magnetic fields in light of EDGES signal}
	\author{Jitesh R. Bhatt}
	\email{jeet@prl.res.in}
	\affiliation{%
		Physical Research Laboratory, Theoretical Physics Division, Ahmedabad 380 009, India}
	\author{Pravin Kumar Natwariya  \orcidA{}\,}
	\email{pravin@prl.res.in}
	\affiliation{%
		Physical Research Laboratory, Theoretical Physics Division, Ahmedabad 380 009, India}
	\affiliation{%
		Department of Physics, Indian Institute of Technology, Gandhinagar, Ahmedabad 382 424, India}  
	\author{Alekha C. Nayak     \orcidB{}\,}
	\email{alekha@prl.res.in}
	\affiliation{%
		Physical Research Laboratory, Theoretical Physics Division, Ahmedabad 380 009, India}
	\author{Arun Kumar Pandey   \orcidC{}\,}
	\email{arunp77@gmail.com}
	\affiliation{%
		Physical Research Laboratory, Theoretical Physics Division, Ahmedabad 380 009, India}
	\date{\today}
\begin{abstract}
{\bf Abstract:}
We have shown that in presence of a cosmic magnetic field the bounds on baryon dark matter cross-section ($\hat \sigma$), dark-matter mass ($m_d$) and values of the magnetic field ($B_0$) can strongly influence each other. This requires to rework the bounds on $\hat \sigma\,$, $m_d$ and $B_0$ which can explain the observed absorption signal by EDGES collaboration. The upper limit on the magnetic field strength can modify in presence of baryon-dark matter interaction cross-section. In the presence of a strong magnetic field, a large baryon-dark matter interaction cross-section is required to balance magnetic heating of gas to explain the EDGES signal as compared to a weak magnetic field. Subsequently, the strong magnetic-fields can even erase the 21~cm signal--this gives an upper bound on the strength of magnetic-fields, dark-matter mass and baryon-dark matter cross-section.  In the special case when $\hat \sigma$=0, one can recover the bound on magnetic field strength calculated in \cite{Minoda:2018gxj}. In this work we find that the allowed range of the primordial magnetic field can increase by three orders of magnitude in comparison with \cite{Minoda:2018gxj}. We get upper bound on the magnetic field strength: $3.48\times10^{-6}$~G for the dark matter mass $\lesssim 10^{-2}$~GeV. 

\end{abstract}
\maketitle
\flushbottom
\clearpage
\section{Introduction}
\label{sec:intro}
Within the standard cold dark matter cosmology  ($\Lambda$CDM), free electrons and protons cool sufficiently after $3\times 10^{5}$ years of big-bang to form neutral hydrogen atoms. At the end of recombination ($z \approx 1100$), matter decouples from the cosmic microwave background (CMB) photons and the temperature reaches around 3000 K. After that, the dark age begins and the Universe becomes  homogeneous. Later during the cosmic dawn era ($15 < z < 35$), over density grow in the matter perturbations and collapse to form the first star and galaxy in the Universe. During the dark ages, residual electrons from the recombination scatter off the baryons to maintain the thermal equilibrium until $z\approx 200$. Subsequently, gas cools adiabatically due to the thermal expansion of the Universe and the gas temperature reaches to $\approx 6.8$ K at $z=17$.

During the cosmic dawn era, the gas temperature is lower than the CMB temperature hence hyperfine transitions in the neutral hydrogen atoms produce 21 cm absorption spectra. The hyperfine transitions are due to the CMB photons, gas collisions and the Ly-$\alpha$ radiations from the first star. 21 cm absorption line leaves an imprint of spectral distortion in the low-frequency tail of the CMB spectrum. The observation of the 21 cm line can give logical reason behind the density fluctuations \cite{Hogan:1979mj}, cosmic reionization \cite{Scott:1990mj} and X-ray heating of the Inter-Galactic Medium (IGM) \cite{Fialkov:2014kta}.  Recently, ``Experiment to Detect the Global Epoch of Reionization Signature" (EDGES) collaboration reported the first detection of such absorption signal centered at 78 MHz \cite{Bowman:2018yin}. The observed absorption dip at $z \approx 17$ is approximately $2.5$ times larger than the standard $\Lambda$CDM model prediction \cite{Bowman:2018yin}. Several attempts have been made in the literature to explain the observed EDGES anomaly.

 There are two ways to explain this enhanced 21 cm absorption signal: one is to the heating of CMB photons and another way is to cool the gas in IGM below the standard $\Lambda$CDM prediction. The first possibility has been investigated by authors of the Ref. \cite{Moroi:2018vci, FRASER2018159,PhysRevLett.121.031103,Liu:2019H}.  In the second scenario, IGM gas can be cooled by emitting the photons between the Ly-limit to Ly-$\gamma$ wavelengths \cite{Chuzhoy_2007,Chuzhoy2006}. There are very few mechanisms to cool the gas. Since the dark-matter is colder than the gas, effective cooling of the gas can be obtained by elastic scattering between the dark-matter and baryon particles \cite{Barkana:2018nd, Barkana:2018lgd, Tashiro:2014tsa}. Therefore, a non-standard Coulomb interaction between dark-matter and baryon: $\sigma= \hat{\sigma} v^{-4}$ can be  considered to explain the EDGES signal, where $v$ is the relative velocity between the dark matter and baryon \cite{Bowman:2018yin,Tashiro:2014tsa, Dvorkin:2013cea,Barkana:2018nd}.  In these  cases, dark-matter considered as millicharge-dark-matter \cite{Bowman:2018yin,Barkana:2018lgd}.  Using this mechanism 21 cm  absorption signal can be explained \cite{Barkana:2018lgd,Barkana:2018nd,Bowman:2018yin,Tashiro:2014tsa}. In these scenarios, cooling of the gas, by transferring energy to the dark-matter, is tightly constrained because of constraints on the dark-matter mass and cross-section by cosmological and astrophysical observations \cite{Barkana:2018nd, Barkana:2018lgd, Berlin:2018sjs, Creque-Sarbinowski:2019mcm}.

It is observed that the magnetic fields (MFs) are present on all length-scales (i.e. at the length-scales of galaxies and the clusters). These MFs have a strength of the order of few  $\mu$G in the intergalactic medium \cite{Kronberg:1994pp, Neronov:1900zz}. The strength of these MFs is constrained from the structure formation, big bang nucleosynthesis (BBN), temperature anisotropies and polarization of CMB \cite{Trivedi:2012ssp, Trivedi:2013wqa,Sethi:2004pe}. Further, individual limits of the order of $\lesssim$~nG on primordial magnetic fields (PMFs) for various cases  has been reported by Planck 2015 results  \cite{Ade:2015cva}. From the data of Fermi and High Energy Stereoscopic System (HESS) gamma-ray telescopes, the peak strength of the magnetic field at a length scale of 1 Mpc can be the order of few nG \cite{Neronov:1900zz}. However, the upper bound on the amplitude of the magnetic field, obtained from BBN is $\sim 10^{-6}$ G at a comoving length scale of $\sim 100$ pc \cite{Cheng:1996vn, Grasso:2000wj}. If strong MFs were present at the time of nucleosynthesis, the abundance of relic $^4$He and other light elements could have different values than we observe today 
\cite{Matese:1969cj, Greenstein:1969}. 
To satisfy the current observational limits on the light element abundances, the strength of these MFs can have a value of the order of $\approx 10^{-7}$~G, at present time \cite{Tashiro:2005ua, Sethi:2004pe}.

In the presence of PMFs, decaying magneto-hydrodynamics effects can heat the IGM gas \cite{Sethi:2004pe,Schleicher:2008aa,Chluba2015}. Hence, the thermal evolution of the gas can modify and it can erase the 21 cm absorption signal \cite{Minoda:2018gxj}. Therefore, magnetic heating of the IGM gas can be constrained by the EDGES signal. Energy deposition into the IGM can be done by ambipolar diffusion. To dissipate the energy into IGM gas, for small-scales, can be done by producing decaying magneto-hydrodynamics turbulence \cite{Sethi:2004pe, Sethi:2008eq}. Heating due to the turbulent-decay starts after recombination when radiative viscosity decreases and velocity perturbations are no longer damped \cite{Sethi:2004pe, Schleicher:2008aa}.
Energy dissipation due to the ambipolar diffusion is important in neutral IGM gas \cite{Sethi:2004pe}, for details see \cite{Shu:1992fh}. At the late time, heating of the gas due to the turbulent-decay decreases, but due to the ambipolar diffusion it continues \cite{Chluba2015, Sethi:2004pe}. In the context of EDGES anomaly, this magnetic heating of the IGM gas has been studied by authors of the Ref. \cite{Minoda:2018gxj}. By the constraint on gas temperature during redshift  $15\lesssim z\lesssim20$ (EDGES result), they put a constraint on the upper limit of the PMFs strength: $B\lesssim 10^{-10}$~Gauss at the length-scale of 1~Mpc.

If one invokes cooling of gas beyond the standard scenario in the presence of baryon dark matter interaction, discussed above, this bound on the strength of PMFs (obtained in the Ref. \cite{Minoda:2018gxj}) may change by transferring energy of the gas to the dark matter (DM) using drag between gas and DM. This transfer of energy depends on the dark-matter mass and cross-section. Therefore, we also get the constraint on DM mass and cross-section to get the 21~cm absorption signal. 

There are also several other ways to heat the gas: Heating of the neutral hydrogen in the IGM due to Ly-$\alpha$ photons has been investigated in some Refs. \cite{Oklop_i__2013,Chuzhoy2007,Chuzhoy_2007,Ghara2019}, annihilation of DM can inject the energy into IGM \cite{PhysRevD.93.023527,PhysRevD.93.023521,PhysRevLett.121.011103,Liu:2018uzy}, DM decay can also heat the IGM gas via energy deposition \cite{Mitridate_2018,PhysRevD.95.023010,Liu:2018uzy}. In Ref. \cite{Bhatt:2019qbq}, authors have considered the dark matter viscosity to heat the gas and dark matter in the context of observed 21 cm signal \cite{Sethi:2004pe, Schleicher:2008aa, Chluba2015}. Here it is to be noted that, we are interested only in the magnetic heating of the gas in the presence of DM-baryon interaction due to the effective cooling of IGM by DM.  The origin of these PMFs fields could be due to some high energy process in the very early universe \cite{turner:1988mw, Sharma:2018kgs, Pandey:2015kaa, Anand:2017zpg, Subramanian:2015lua}.

The present work is divided into the following sections: in section (\ref{sec:21cm}), we have revisited the 21 cm observed signal from the EDGES; in section (\ref{sec:PMFrecom}), a brief description of the PMFs and it's decay is discussed; section (\ref{gas_dm_mag}) contains the result obtained and a detailed discussion.
In the end, we have summarized and concluded  our work in section (\ref{sec:conc}).
\section{21 cm signal and EDGES observation} 
\label{sec:21cm}
At the end of recombination, the baryon number density of the Universe mostly dominated by the neutral hydrogen, small fraction of helium, residual free electrons and protons. The hyperfine interaction between spin of the electron and proton split the ground state of neutral hydrogen atom into singlet and triplet states with an energy difference of $E_{21}=5.9 \times 10^{-6}$~eV $=2 \pi/(21 ~ \text{cm})$. Relative number densities of neutral hydrogen in the singlet and triplet state, define the spin temperature ($T_S$) of the gas, and is given by the following relation:
\begin{equation}
	\frac{n_{1}}{n_{0}} = \frac{g_{1}}{g_{0}}~e^{-\frac{ E_{21}}{T_{S}}} \simeq 3\left( 1 - \frac{ E_{21}}{T_{S}}\right),
\end{equation}
here, $n_0$ and $n_1$ are the number densities in the singlet and triplet states respectively. $g_0$ and $g_1$ are statistical
degeneracy of singlet and triplet states respectively. Spin temperature depends on gas collision,  emission/absorption of CMB photon and Ly-$\alpha$ radiation from the first star. Equilibrium balance between the populations of singlet and triplet state describes the spin temperature \cite{Field,Pritchard:2011xb}

\begin{equation}
	T_{S}^{-1} = \frac{T_{\rm CMB}^{-1} + x_{c}T_{\rm gas}^{-1} + x_{\alpha}T_{\rm Ly\alpha}^{-1}}{1 + x_{c} + x_{\alpha} }
	\label{eq:spintemp}\,,
\end{equation}

where, $T_{\rm CMB}$ is the CMB temperature, $T_{\rm gas}$ is the kinetic temperature of the gas and $T_{\rm Ly\alpha}$ is the colour temperature of Ly$\alpha$ radiation from the first star. Here, $x_c$ and $x_{\alpha}$ are the collisional and Ly$\alpha$ coupling respectively \cite{Field}, 
\begin{equation}
	x_{c} = \frac{E_{21}}{T_{\rm CMB}}\frac{C_{10} }{A_{10} }\ , \ \  x_{\alpha} = \frac{E_{21}}{T_{\rm CMB}}\frac{P^{w}_{01} }{A_{10} }\ , \nonumber
\end{equation}
 where, $C_{10}$ is the collision rate between $H-H,~H-e,~p-H$ and $P_{01}^{w}$ is the excitation rate due to Ly-$\alpha$ radiation and $A_{10}=2.9\times 10^{-15}$sec$^{-1}$ is the Einstein coefficient for spontaneous emission. After the first star formation, a large number of Ly$\alpha$ photons scattered with the gas, and brought the radiation and the gas into a local thermal equilibrium \cite{Pritchard:2011xb}. Hence during the cosmic dawn era $T_{\rm gas} \approx T_{\rm Ly\alpha}$. The 21 cm signal can be described, in terms of the redshifted differential brightness temperature \cite{Pritchard:2011xb} 
\begin{equation}
	T_{21}= \frac{1}{1+z}(T_{S}-T_{\rm CMB})(1-\exp^{-\tau})
	\label{eq:bright}\,,
\end{equation}
where, optical depth $\tau=\frac{3 \lambda_{21}^2A_{10}n_H}{16 T_S H(z)}$, $n_H$ is the hydrogen number density, $\lambda_{21}=21$~cm and $H(z)$ is the Hubble rate. Depending on the spin and the CMB temperature, three scenarios arise: (i) when $T_S~=T_{\rm CMB}$, then no signal is observed; (ii) if $T_S< T_{\rm CMB}$, photon get absorbed by the gas and absorption spectra is observed; and (iii) if $T_S> T_{\rm CMB}$, then it leaves an imprint of emission spectra. 

Evolution of the 21 cm signal is as follows: after recombination at $z \sim 1100$ down to $z \sim 200$, the residual free electrons undergo Compton scattering to maintain thermal equilibrium between the gas and CMB, and collisions among the gas is dominant, i.e. $x_c \gg 1,x_{\alpha}$ \cite{Barkana:2018nd,Pritchard:2011xb}, which set $T_S=T_{\rm CMB}$. Hence, 21 cm signal is not observed during this era. Below $z \sim 200$ until $z \sim 40$, the gas cools adiabatically and it falls below CMB temperature, which implies the early 21 cm absorption signal. The sensitivity of radio
antennas below 50~Mhz falls dramatically and collisional absorption signal can not be observed. Below $z\sim 40 $ down to the first star formation, gas cools sufficiently due to the expansion and the collisional coupling becomes very small due to the dilution, i.e. $x_{c},~ x_{\alpha} \rightarrow 0$ \cite{Barkana:2018nd,Pritchard:2011xb}. This implies, $T_S= T_{\rm CMB}$, hence no $21$ cm signal during this period.  After the first star formation, a transition between the singlet and triplet states occurs due to the Ly$\alpha$ photons emitted from the first stars via Wouthuysen-Field (WF) effect \cite{1952AJ.....57R..31W,1959ApJ...129..536F}.  Ly$\alpha$ photons couple the spin temperature to the gas temperature. In this era, $x_{\alpha} \gg  1$, hence spin temperature and the gas temperature become equal to each other i.e., $T_S=T_{\rm gas}< T_{\rm CMB}$. Thus an imprint of the 21 cm absorption signal can be seen at the low-frequency tail of the CMB spectrum. Below $z\sim 15$ until $z \sim 7$, X-ray from the active galactic nuclei heats the gas above the CMB temperature and we observe an emission signal \cite{Pritchard:2011xb}. Below $z\sim 7 $ neutral hydrogen became ionized and the signal disappears. 

Recently, the EDGES collaboration reported the global brightness temperature at $z=17$ 
\begin{equation}
T_{21}^{\rm obs}(z=17)= -500^{+200}_{-500}~ {\rm mK}\ , 
\end{equation}
and corresponding gas temperature is $3.26^{+1.94}_{-1.58}$ K \cite{Bowman:2018yin}. On the contrary, standard $\Lambda$CDM predicts the gas temperature 6.8 K at $z=17$ and corresponding brightness temperature $T_{21} \geq -220$ mK \cite{Barkana:2018nd}. In order to explain the EDGES absorption signal, the gas temperature needs to be cooler than the $\Lambda$CDM prediction. During the Cosmic dawn era, the Universe was at its coldest phase, and the relative velocity between the DM and baryon was very small [$\mathcal{O}(10^{-6})$]. Also, the temperature of the dark matter is colder than the baryon temperature during this period, so an interaction of the baryon with dark matter can cool the gas temperature. Since the relative velocity is small, scattering cross section of the type $\sigma= \hat{\sigma} v^{-4}$ can enhance the interaction rate and cool the gas sufficiently to explain EDGES absorption dip \cite{Tashiro:2014tsa, Dvorkin:2013cea}. In this work, we consider magnetic heating of the gas and the DM via ambipolar and turbulent decay.
\section{Primordial magnetic fields after the recombination era}
\label{sec:PMFrecom}

In this section, we  present the evolution of PMFs via two processes i.e. ambipolar diffusion and decaying turbulent. We assume that, due to some early Universe process, tangled magnetic fields were present at a sufficiently large length scale after the recombination era \cite{turner:1988mw, Sharma:2018kgs, Pandey:2015kaa, Anand:2017zpg, Subramanian:2015lua}. 
 We also consider a very small velocity induced by the PMFs to avoid any dissipations of the initial magnetic fields due to viscosity and other dissipation \cite{Jedamzik:1998kk, Subramanian:1997gi}. This is applicable in linear regime and magnetic fields evolve adiabatically as ${\bf B}(t, {\bf x})=\tilde{{\bf B}}({\bf x})/a(t)^2$, where ${\bf x}$ is the comoving coordinate, $\tilde{{\bf B}}$ is the comoving strength of the magnetic fields and $a(t)$ is the scale factor. Since plasma in the early Universe remains highly conductive, the adiabatic evolution of the magnetic fields is true. However, at a sufficiently small scale, when non-linear effects operate, adiabatic decay no longer satisfy. In this case, it is needed to consider Euler Eqs. along with the magnetic induction equation and Maxwell's Eqs. to understand the dynamics of the fluid.
We consider an isotropic and homogeneous Gaussian random magnetic field, whose power spectrum is given by the following equation
\begin{equation}
\langle \tilde{{\bf B}}_i ({\bf k}) \, \tilde{{\bf B}}^*_j ({\bf q})\rangle = \frac{(2\pi)^3}{2}\delta_D^3({\bf k}-{\bf q})\left(\delta_{ij}-\frac{k_i k_j}{k^2}\right)\mathcal{P}_B(k)\,,
\end{equation}
where, $\mathcal{P}_B(k)$ is the magnetic power spectrum and $k=|{\bf k}|$ is the comoving wave number. For simplicity, we consider a power law spectrum of the magnetic fields $\mathcal{P}_B(k)= A\, k^{n_B}$ for $k< k_{\rm max}$ ($k_{\rm max}$ is calculated by the damping at recombination due to the viscosity) \cite{Jedamzik:1998kk, Subramanian:1997gi}. Here $n_B$ and $A$ are spectral index and the normalization constants respectively. In particular, $n_B=2$ for white noise \cite{Hogan:1983cj}, $n_B=4$ for Batchelor spectrum \cite{Durrer:2003ja} and $n_B=-2.9$ for nearly scale invariant spectrum \cite{Sethi:2004pe}. The amplitude ($A$) can be obtained by demanding the magnetic fields are smooth over the cut off scale, and after that $\mathcal{P}_B(k)=0$. Once the recombination period end, baryons and CMB photons decouple and their velocity  start to increase. Eventually, these particles achieve a common velocity, determined by the equipartition between the magnetic field and kinetic energy of the baryon gas. This velocity is given by the Alf\'ven velocity $v_A=c\, B_0/\sqrt{4\pi \rho_{b_0} a(t)}$, here $\rho_{b_0}$ is the present baryon density and $B_0$ is the currently observed strength of the magnetic field at Mpc scale. The cut off scale is defined as $k_{{\rm max}}\approx 2\pi \frac{H a}{v_A}$. Therefore, the cut off value from this relation can be written as \cite{Sethi:2004pe, Tashiro:2006uv}:
\begin{equation}
\frac{k_{\rm max}}{2 \pi \,{\rm Mpc}^{-1}} = \left[1.32\times 10^{-3} \left(\frac{B_0}{1 {\rm nG}}\right)^2 \left(\frac{\Omega_b h^2}{0.02}\right)^{-1}\left(\frac{\Omega_m h^2}{0.15}\right)\right]^{\frac{-1}{(n_B+5)}}\,,
\end{equation}
here, $\Omega_b$, $\Omega_m$, and $h$ are the cosmological parameters and have the mathematical value $h=0.674$ ($H_0=100\,h$~Km/s/Mpc), $h^2\Omega_b=0.0224\pm 0.0001$, $h^2\Omega_m=0.143$ \cite{Aghanim:2018eyx}.
\subsection*{Dissipation of magnetic energy}
 After recombination, any present magnetic field dissipates its energy and heats the gas through two mechanisms, namely the ambipolar diffusion and the turbulent decay \cite{Sethi:2004pe, Tashiro:2006uv}. The heating of the gas gives a considerable change in the thermal evolution of neutral atoms. The velocity difference in the ionized and neutral particles after the recombination leads to the ambipolar diffusion of the magnetic energy. Direct cascade happens due to the non-linear processes, which couples the different modes and the cascading of magnetic energy from large to small scale. This happens through the  breaking  of the larger eddy into the smaller eddies, when eddy turn over time $t_{\rm eddy}$ is equal to the Hubble time i.e. ($t_{\rm eddy}\sim H^{-1}$). The energy dissipation due to ambipolar and direct cascade can be given by \cite{Cowling:1956gt, Shu:1992fh},
\begin{eqnarray}
\Gamma_{{\rm ambi}} & = & \frac{\rho_n}{16\pi^2\gamma \rho_b^2 \rho_i} |(\nabla\times {\bf B})\times {\bf B}|^2\,, \label{eq:ambi1} \\
\Gamma_{{\rm decay}} & = & \frac{B_0^2(t)}{8\pi}\frac{3m}{2}\frac{\left[\ln\left(1+t_{\rm eddy}/t_i\right)\right]^m\, H(t)}{\left[\ln\left(1+t_{\rm eddy}/t_i\right)+\ln(t/t_i)\right]^{m+1}}\,, \label{eq:decay1}
\end{eqnarray}
where, $\rho_n$ and $\rho_i$ are the mass densities of neutral and the ionized atoms respectively, $t$ is the cosmological time at a generic red shift $z$, $t_{\rm eddy}$ is the physical decay time scale for the turbulent and $t_i$ is initial time at which decay starts.  $m= 2(n_B+3)/(n_B+5)$. For the present scenario $\gamma$ is given by \cite{Shu:1992fh, Shang:2001df, Draine:1980bt, Schleicher:2008aa},
\begin{equation}
\gamma =\frac{\frac{1}{2}n_H\langle\sigma v\rangle_{{\rm H}^+, {\rm H}}+\frac{4}{5}n_{{\rm He}}\langle\sigma v\rangle_{{\rm H}^+,{\rm He}}}{m_H[n_H+4n_{{\rm He}}]}\,,
\end{equation}
where, $m_H$ is the mass of the hydrogen and $n_{\rm He}$ is the number density of the helium atom. The absolute values of the Lorentz force and magnetic energy in Eqs. (\ref{eq:ambi1}) and (\ref{eq:decay1}) can be obtained by considering a suitable power law spectrum of the magnetic field. This can be done using following correlation integrals: $|(\nabla\times {\bf B})\times {\bf B}|^2= \int (dk/2\pi)^3\int (dq/2\pi)^3\, k^2\, \mathcal{P_B}(t, k)\mathcal{P_B}(t, q)(1+z)^{10}$ and $|{\bf B}|^2 =\int (dk/2\pi)^3 \mathcal{P_B}(t, k) (1+z)^4$. However, in the present work, we have taken approximate value of the ambipolar diffusion term, which is \cite{Schleicher:2008hc}
\begin{equation}
\Gamma_{{\rm ambi}}\sim \frac{\rho_n}{16\pi^2\gamma \rho_b^2\rho_i}\, \frac{B^4}{L^2}\,,
\end{equation}
here, $L$ is a typical length scale. To sum up, the time evolution of the magnetic energy can be written as \cite{Subramanian:1997gi, Sethi:2004pe}
\begin{equation}
\frac{d}{dt}\left(\frac{|{\bf B}|^2}{8\pi}\right)= -4 H(t)\left(\frac{|{\bf B}|^2}{8\pi}\right)-\Gamma_{{\rm ambi}}-\Gamma_{{\rm decay}}\,.
\label{eq:mag_heat}
\end{equation}
%
\section{Baryon and dark matter interaction in presence of magnetic field}
\label{gas_dm_mag}
In this section, we discuss the effects of magnetic field on the baryon and DM temperature when they are interacting with each other. Temperature evolutions of the DM and baryon, having relative velocity, in the presence of the magnetic field are given below
\begin{eqnarray}
\frac{dT_{\rm gas}}{dz} & = & \frac{2T_{\rm gas}}{(1+z)} + \frac{\Gamma_{C}}{(1+z)H} (T_{\rm gas}-T_{\rm CMB}) \nonumber \\ 
&-&  \frac{2}{3(1+z)H}\frac{d{Q_{gas}}}{dt}-  \frac{2\Gamma_{\rm heat}}{3n_b(1+z)H}\,,
\label{eq:baryon_temp} \\
\frac{dT_{d}}{dz}& = &\frac{2T_{d}}{(1+z)} -  \frac{2}{3(1+z)H}\frac{d{Q_{d}}}{dt}
\label{eq:dm_temp}\,, \\
\frac{dv}{dz}& =&  \frac{v}{(1+z)} + \frac{D(v)}{(1+z)H}\,,
\label{eq:vel}
\end{eqnarray}
where,  $T_{d}$ and $m_d$ are temperature and  mass of the DM respectively, $H$ is the Hubble expansion rate, $n_b$ is the baryon number density and $\Gamma_C$ is the Compton scattering rate, defined as
\begin{eqnarray}
\Gamma_{C}= \frac{8 \sigma_T a_r T_{\rm CMB}^4 x_e}{3\,(1+x_e+x_{He})m_e}\,.\nonumber
\end{eqnarray}
Here, $x_e=n_e/n_H$ is the electron fraction, $x_{He}$ is the helium fraction, $a_r$ is the Stefan-Boltzmann radiation constant and $\sigma_T$ is the Thomson scattering cross section.  Drag term  $D(v)$ is   given by
\begin{equation}
D(v) \equiv  \frac{\rho_{m}\hat{\sigma}}{m_{H} + m_{d}}\frac{1}{v^{2}}F(r) \label{eq:drag1}\ ,
\end{equation}
where, $ \rho_{M} $ is  matter density and 
\begin{equation}
F(r) \equiv \mathrm{erf}(\frac{r}{\sqrt{2}}) - \sqrt{\frac{2}{\pi}}r e^{-r^{2}/2} \nonumber\ ,
\end{equation}
with $r = \frac{v}{u_{\mathrm{th} }}$ and $u_{\mathrm{th} } =\sqrt{ \frac{T_{\rm gas}}{m_{H}} + \frac{T_{d}}{m_{d}}}~. $  Heat transfer of the gas per unit time is given by \cite{Munoz:2015bk}
\begin{equation}
\frac{d{Q_{gas}}}{dt} = \frac{2 m_H\rho_{d}\hat{\sigma}e^{-r^2/2}}{\sqrt{2 \pi}(m_{H} + m_{d})^2 u_{\mathrm{th}}^3} \bigg( T_{\rm d} - T_{gas}\bigg)
+\frac{\mu \rho_{d} }{\rho_{M}} v D(v)~,
\label{eq:heat_tranfer}
\end{equation}
where,  $\mu$ is the reduced mass of DM and baryon, $\hat{\sigma}$ is  DM-gas scattering cross-section and $\rho_d$ is
the DM energy density. In Eq.\eqref{eq:heat_tranfer}, the first term  represents the baryon's cooling due to its interaction with the DM and second term represents the heating due to drag term. Relative velocity between DM and gas generates friction between two fluid which is responsible for the drag term. Heat transfer rate for dark matter $\left(\frac{d{Q_{d}}}{dt}\right)$ can be obtained by interchanging $gas \leftrightarrow d$ in Eq. \eqref{eq:heat_tranfer}.
Temperature evolution of the DM and gas require electron ionization fraction \cite{AliHaimoud:2010dx}:
\begin{eqnarray}
\frac{dx_e}{dz} & = & \frac{1}{H(1+z)} \frac{\frac{3}{4}R_{Ly\alpha} + \frac{1}{4} \Lambda_{2s,1s}}{\beta_{B}+\frac{3}{4}R_{Ly\alpha}+\frac{1}{4}\Lambda_{2s,1s}}\nonumber \\
& &\times\,\Big(n_H x_e^2\alpha_{B}-4(1-x_e)\beta_{B}e^{-E_{21}/T_{\rm CMB}} \Big)\,,
\label{electron}
\end{eqnarray}
where, $\beta_{B}$ and $\alpha_{B}$ are the photo-ionization rate and case-B recombination coefficient respectively.  $E_{21}$ is energy of Ly$\alpha$ wavelength photon and $\Lambda_{2s,1s}=8.22~ {\rm sec}^{-1}$ is the  hydrogen two photon decay rate. The  escape rate of  Ly$\alpha$  is given by: $R_{\text{Ly}\alpha}=\frac{8 \pi H}{3 n_H (1-x_e)\lambda_{\text{Ly}\alpha}^3}$, $\lambda_{\rm Ly\alpha}$ is the rest wavelength of Ly$\alpha$ photon. As it has been confirmed in Ref. \cite{Minoda:2018gxj}, that cooling due to effects like Lyman-$\alpha$ emission, Bremsstrahlung and recombination does not have that much effects on the dynamics of the gas and DM, therefore, we have not considered these effects in our present work.
\begin{figure*}[t]
	\subfloat[]{\includegraphics[width=0.35\linewidth, keepaspectratio]{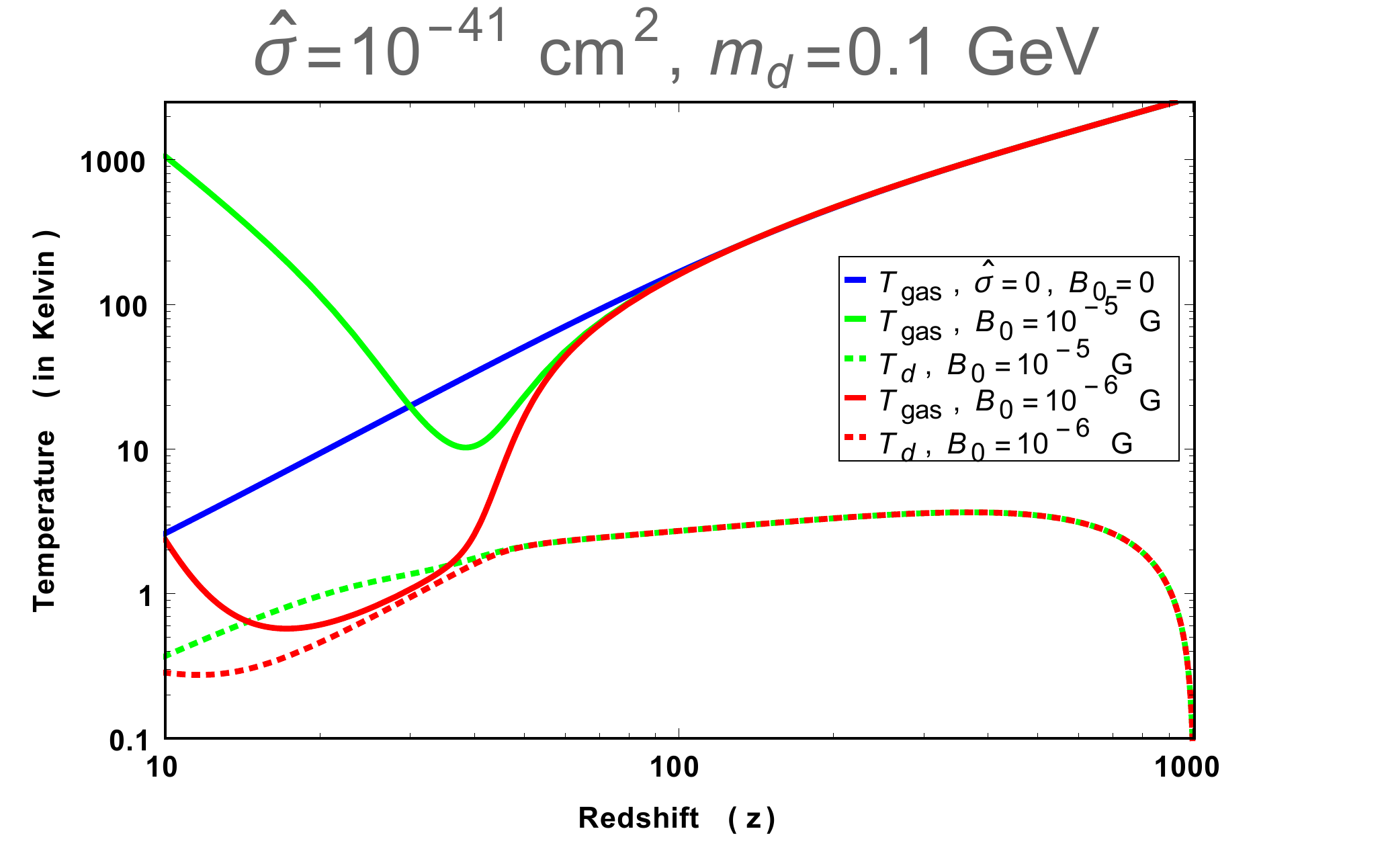}\label{fig:Temp-crossmass}}
	\subfloat[]{\includegraphics[width=0.35\linewidth, keepaspectratio]{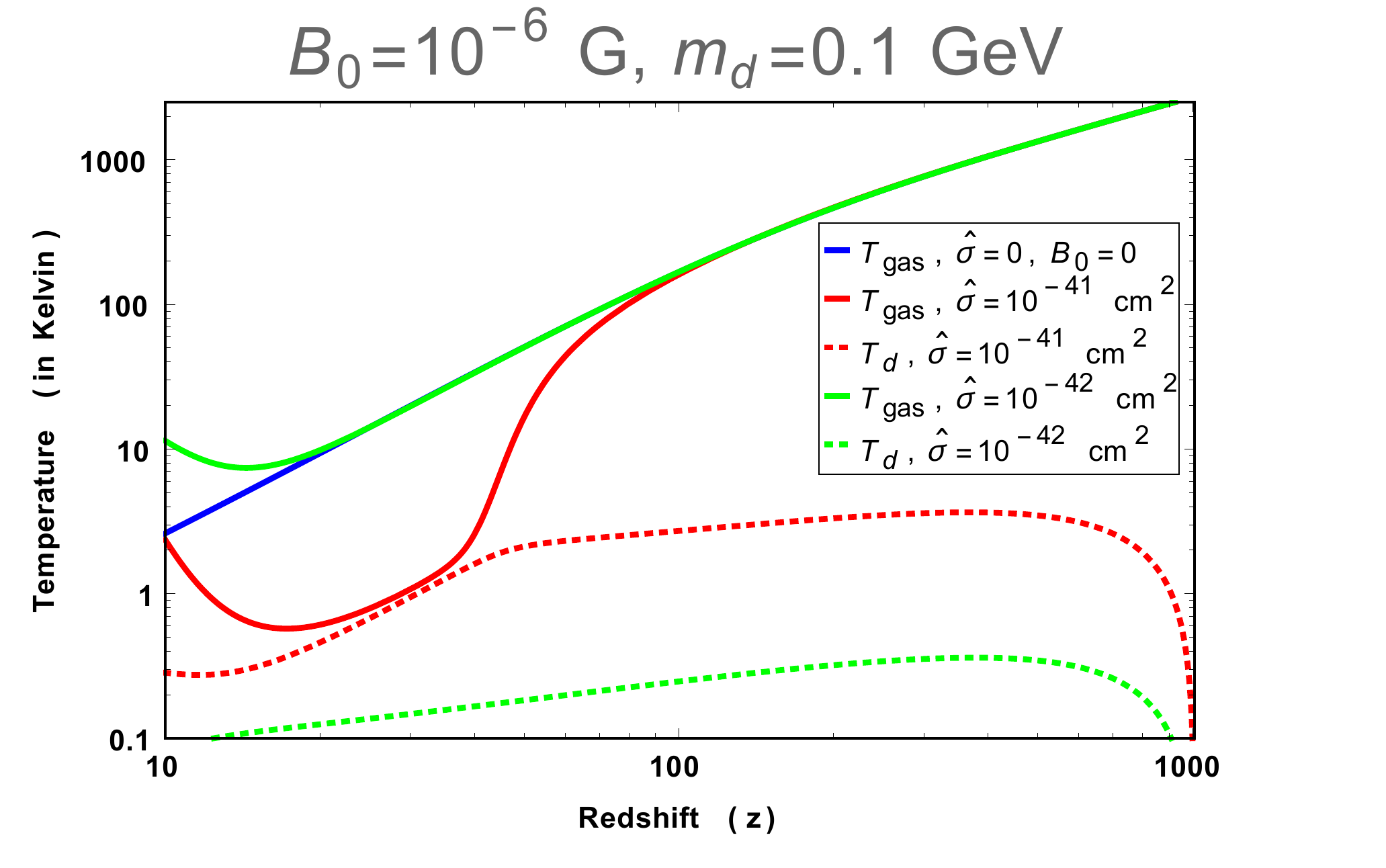}\label{fig:Temp-cross-mass}} 
	\subfloat[]{\includegraphics[width=0.35\linewidth, keepaspectratio]{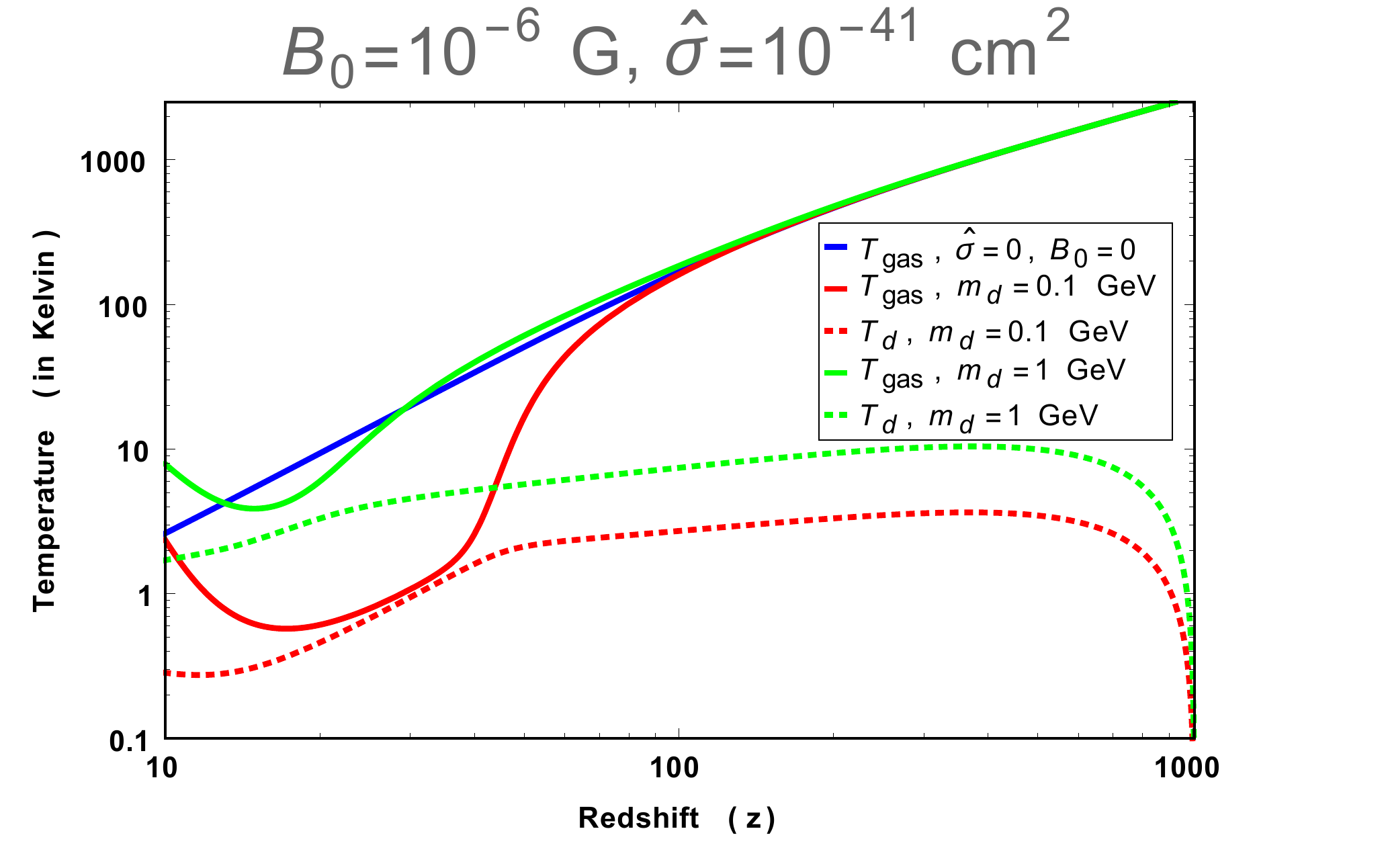}\label{fig:Tem-B-cross}}
\caption{\raggedright Temperature evolution of baryon and DM in the presence of magnetic field and baryon-DM interaction. Blue
	line corresponds to temperature evolution of gas in the absence of both magnetic heating and baryon-DM
	interaction. The red (green) solid lines represents the variation of the gas temperature and the dotted red (green)
	line shows the variation of the DM temperature in presence of magnetic field and the baryon-DM interaction.}
\label{fig:detail}
\end{figure*}
\begin{figure}[ht]
    \includegraphics[width = 3.5in, ,height=2.1in]{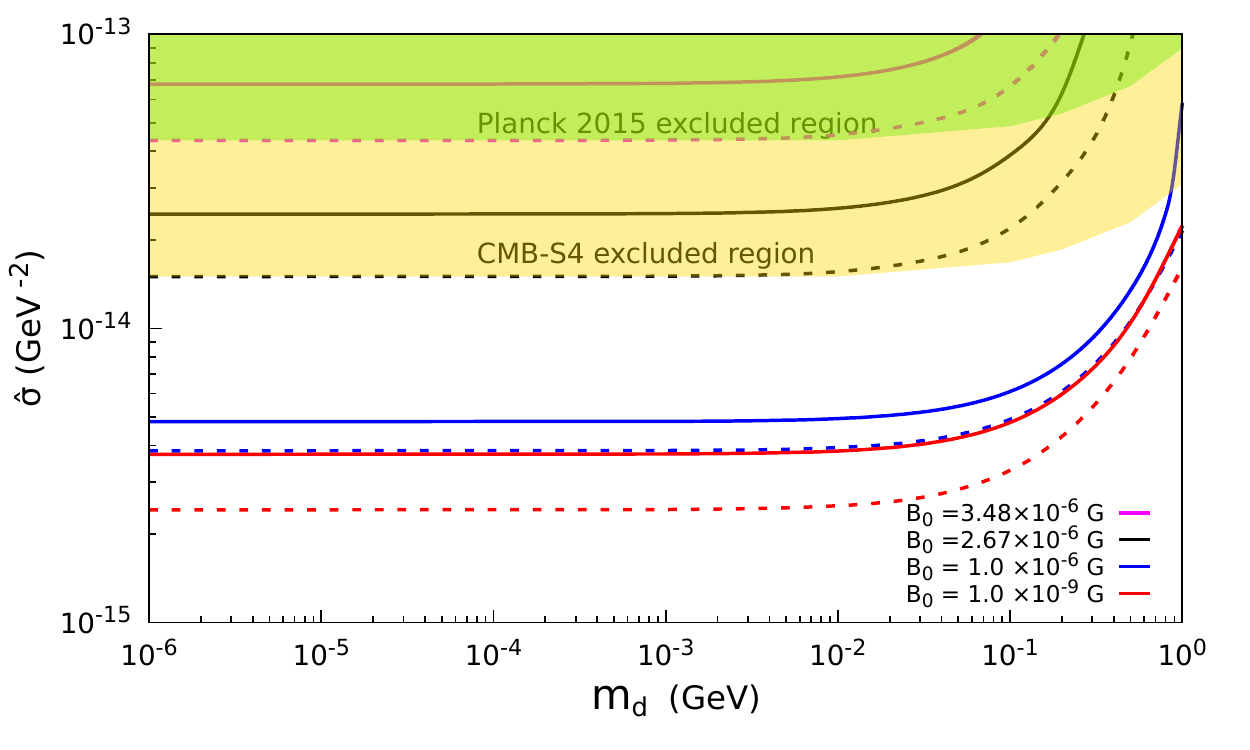}
    \caption{ \raggedright Constrains on $\hat \sigma$ and $m_d$ for different magnetic field strengths by requiring $T_{21}\simeq-500$~mK (solid lines) and $T_{21}\simeq-300$~mK (dashed line) at $z=17$. The solid (dashed) magenta, black, blue and red line correspond to $B_0=3.48\times10^{-6}$~G, $2.67\times10^{-6}$~G, $10^{-6}$~G and $10^{-9}$~G respectively. The CMB-S4 (forecast) and Planck 2015 constraints on $\hat \sigma$ and $m_d$ with 95\% C.L. have been taken from the Refs. \cite{Kovetz:2018P, Boddy:2018G}. The green and gold regions are excluded by Planck 2015 and CMB-S4 forecast respectively.   ($1$ GeV$^{-2}=3.89 \times 10^{-28}$ cm$^2$) }\label{fig:cross-mass}
\end{figure}
\section{Results and Discussion}\label{sec:res-dis}
Solving Eqs. (\ref{eq:mag_heat}-\ref{eq:vel})  and Eq. (\ref{electron}) with initial conditions $T_{\rm gas}(1010)=T_{\rm CMB}(1010)=2749.92$~K, $T_d(1010)\sim 0$~K, $x_e(1010)=0.057$ and $B(z)=B_0\, (1+z)^2|_{z=1010}$ initial magnetic field strength, we get the temperature evolution of the DM and gas for different DM masses, DM-baryon interaction cross-sections and MF's strengths. Figure \eqref{fig:detail} shows the evolution of the gas and DM temperature with redshift ($z$). The solid blue line in Fig. \eqref{fig:detail} corresponds to gas temperature when both the magnetic field and DM-baryon interaction are zero. In this case, gas temperature falls as $T_{\rm gas} \propto (1+z)^2$ and reaches 6.8 K at $z=17$. In figure \eqref{fig:Temp-crossmass}, temperature evolution of the gas and DM  is given for different MFs at constant  $\hat{\sigma} = 10^{-41} ~ {\rm cm}^2$ and $m_d=0.1$~GeV. For $B_0 = 10^{-5} ~ (10^{-6})$ G, gas temperature falls down due to Hubble expansion and DM-baryon interaction till $z \approx 30 ~ (z \approx 20)$, then temperature rises due to magnetic heating. For $B_0= 10^{-5}$ G, temperature of the DM also increases due to the coupling between DM and baryons at lower redshift. Larger the strength of MFs, earlier the heating begins. Although dark matter temperature at $z \sim 1010$ is taken to be zero, it heats up due to the heat transfer from baryons to DM. By increasing $B_0$, magnetic-heating of the gas rises. Thus, DM temperature grows due to the drag term in Eq. \eqref{eq:drag1} and it can be seen in Fig. \eqref{fig:detail}, temperature of DM for $B_0=10^{-5}$~G is larger compared to $B_0=10^{-6}$~G. Figure \eqref{fig:Temp-cross-mass} shows the temperature evolution of gas and DM for different DM-baryon interaction cross-section when magnetic field $B_0=10^{-6}$ G and DM mass $m_d=0.1$ GeV are fixed. Larger the $\hat\sigma$ between gas and DM, more heat transfers from gas to the DM and cools the gas efficiently. For $B_0=10^{-6}$ G and $\hat{\sigma} = 10^{-41} ~{\rm cm}^2$, temperature evolution for different dark matter mass is shown in Fig. \eqref{fig:Tem-B-cross}. As we increase the DM mass from $0.1$ GeV to $1$ GeV, temperature of both the DM and gas increases and becomes more efficient for large dark matter mass \cite{Munoz:2015bk}.  This drag heating is important when mass of DM is comparable or greater than 1 GeV \cite{Munoz:2015bk}. Below $z \sim 50$, in addition to heating due to the drag term, magnetic heating also contribute to the gas temperature, hence the  gas temperature for $m_d=1 $~GeV is higher than $m_d=0.1$~GeV.
%
\begin{figure*}[t]
	\centering
	\subfloat[] {\includegraphics[width=3.5in,height=2.0in]{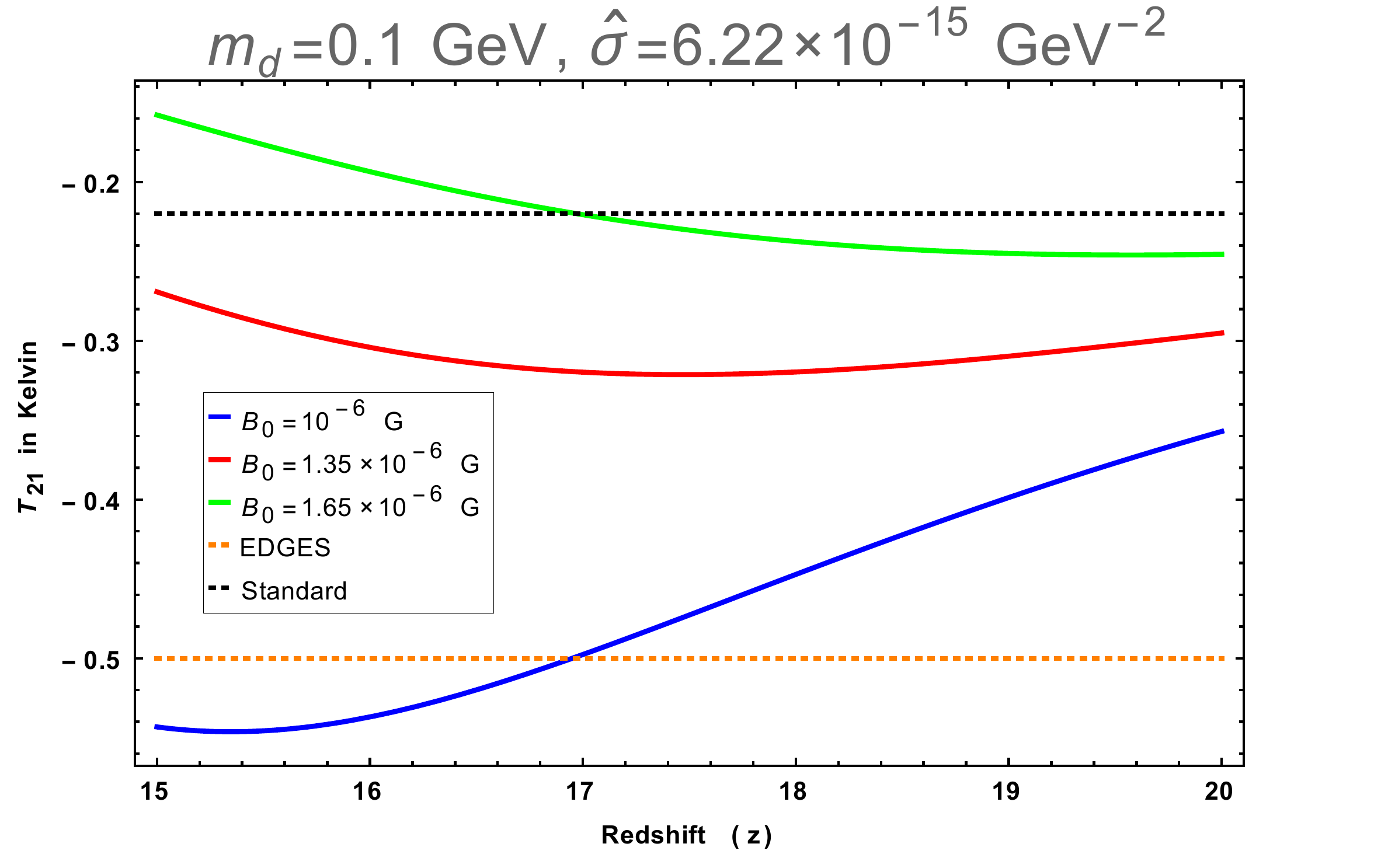}\label{plot:1a}}
	\subfloat[] {\includegraphics[width=3.5in,height=1.9in]{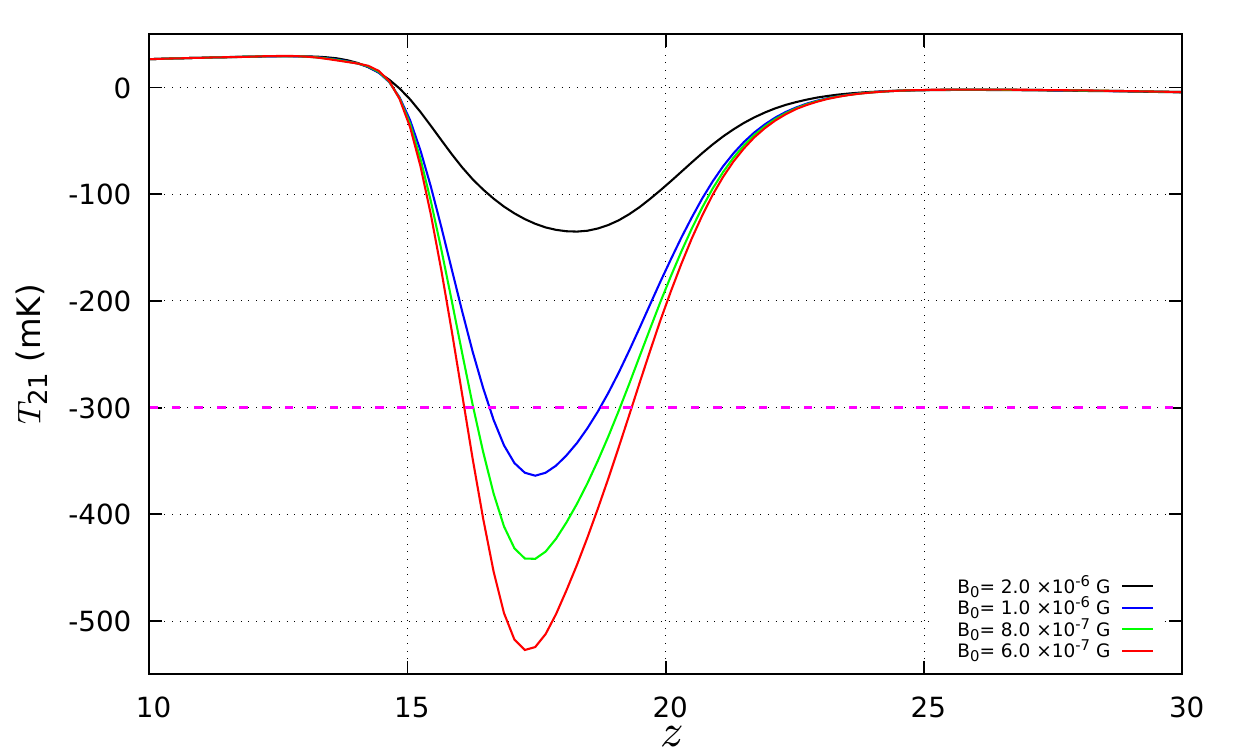}\label{plot:1b}} 
	\caption{ (1a): 21 cm differential brightness temperature (assuming infinite Ly$\alpha$ coupling) vs redshift when their is no X-ray heating.  The dotted black (orange) color represents standard $\Lambda$CDM (EDGES) predictions for the global $T_{21}$ signal. Green, red and blue solid curves correspond to $B_0= 1\times 10^{-6}$, $1.35 \times 10^{-6}$ and $ 1.65\times 10^{-6}$~G respectively. (1b): $T_{21}$ plot with redshift when X-ray heating is included. Black, blue, green and red solid curves correspond to $B_0= 2\times 10^{-6}$, $1 \times 10^{-6}$, $ 8\times 10^{-7}$ and $ 6\times 10^{-7}$~G respectively. The magenta dashed line is  corresponds to the EDGES upper bound on $T_{21}: -300$~mK. For both cases, $m_d=0.1$ GeV and $\hat{\sigma}=6.22\times10^{-15}$~GeV$^{-2}$. }
	\label{fig:mag_21}
\end{figure*}    
\subsection*{Correlation between mass of DM and baryon-DM cross section from EDGES observation}\label{dm-cross}
 In this subsection, we analyze the effect of  $B_0$,  $m_d$ and  $\hat\sigma$ on gas and dark-matter temperature. In  Fig. \eqref{fig:cross-mass}, we study constraints on  $m_d$ and $\hat\sigma$ for  $T_{21}\simeq-500$~mK ($T_{\rm gas}\simeq3.26$~K) and  $-300$~mK ($T_{\rm gas}\simeq5.2$~K).   Here, to calculate $T_{21}$ we have taken $x_\alpha\gg1$. Thus, from equation \eqref{eq:spintemp}, $T_S\approx T_{\rm gas}$ and one can calculate $T_{21}$ from equation \eqref{eq:bright}. Here note that, we do not include heating due to the Ly$\alpha$ background and the effect  of this additional heating become significant  for redshift $z\lesssim$17 for the fiducial model considered in Refs.  \cite{Kovetz:2018P,Harker:2015M,Mirocha:2015G}. We discuss this point in the next subsection. In  Fig. \eqref{fig:cross-mass}, we consider cases $B_0=3.48\times10^{-6}$~G, $2.67\times10^{-6}$~G, $10^{-6}$~G and $10^{-9}$~G and  solve equations (\ref{eq:mag_heat}-\ref{electron}) for $T_{\rm gas}\simeq3.26$ and $5.2$~K at $z=17$ to get $m_d$ vs $\hat{\sigma}$ plots.  The solid and dashed lines represent the cases when $T_{21}\simeq-500$~mK and $-300$~mK respectively. The  gold and green regions respectively show the CMB-S4 (forecast) and Planck 2015 upper constraint on $\hat\sigma-m_d$ with 95\% C.L.  \cite{Kovetz:2018P, Boddy:2018G}. The magenta, black, blue and red lines corresponds to $B_0=3.48\times10^{-6}$~G, $2.67\times10^{-6}$~G, $10^{-6}$~G and $10^{-9}$~G. As we increase the magnetic field strength from $10^{-9}$~G to $\sim10^{-6}$~G, larger value of $\hat\sigma$ is required for $m_d \in \{10^{-6},~ 1\}$~GeV to maintain $T_{21}\simeq-500$ or $-300$~mK at z=17. To get EDGES upper limit on $T_{21}$ (i.e. $-300$~mK), required $\hat{\sigma}$ is smaller compared to the case when $T_{21}=-500$~mK. This is  because we need to transfer less energy from gas to the DM to obtain EDGES upper limit on $T_{21}$. We get the upper limit on magnetic field strength $2.67\times 10^{-6}$~G by CMB-S4 (forecast) constraint on $\hat{\sigma}-m_d$ and maintaining $T_{21}\simeq-300$~mK at z=17. For $B_0=2.67\times 10^{-6}$~G, $m_d \gtrsim 10^{-2}$~GeV is excluded. By Planck 2015 constraint on $\hat{\sigma}-m_d$, the allowed maximum strength of magnetic field is $3.48\times10^{-6}$~G by requiring EDGES upper constraint on $T_{21}$ at z=17. For $B_0=3.48\times10^{-6}$~G, mass of dark-matter $\gtrsim 1\times10^{-2}$~GeV is excluded. Similarly, for the $B_0=10^{-6}$~G, $m_d\gtrsim 8\times10^{-1}$~GeV is excluded by CMB-S4 forecast. When the DM mass approaches mass of hydrogen, the drag term in equation (\ref{eq:drag1}) also starts to contribute in heating of the gas in addition to the magnetic heating. Therefore, higher mass of dark matter is excluded for higher magnetic field. As discussed in \cite{Munoz:2015bk}, when $m_d\gg 1$ GeV, the drag term heat up both the gas and DM in  such a way that we can not obtain $T_{\rm gas}= 3.26$ K at $z=17$ as required for the EDGES signal. There is a independent bound on the primordial magnetic field from CMB of the order of $\lesssim$~nG \cite{Trivedi:2012ssp, Trivedi:2013wqa}. This constraint, in our analysis, restricts value of $\hat\sigma$. Here, we note that in our analysis further decreasing value of $B_0$ below $10^{-9}$~G, does not change our result in significant way.


\subsection*{Effect of strong magnetic field on brightness temperature}{\label{B21}}
We have discussed above that, as we increase the strength of the magnetic field, for a fix DM mass and interaction cross section, temperature of the gas increases. In figure \eqref{fig:mag_21}, we plot 21 cm differential brightness temperature with redshift for different magnetic field strengths. This figure is obtained by keeping $m_d=0.1$~GeV and $\hat{\sigma}=6.22\times 10^{-15}$~GeV$^{-2}$ constant. In figure \eqref{plot:1a}, to plot $T_{21}$ we assume infinite Ly$\alpha$ coupling ($x_\alpha\rightarrow\infty\Rightarrow T_S\simeq T_{\rm gas}$) and do not include the X-ray heating.  For $B_0=1\times10^{-6}$~G, the 21 cm line absorption signal reported by EDGES (i.e. $-500$~mK) can be explained. 
In figure  \eqref{plot:1b}, we include the X-ray heating and consider finite Ly$\alpha$ coupling ($x_\alpha$) \cite{Kovetz:2018P,Harker:2015M,Mirocha:2015G,Zygelman:2005}. As we decrease $B_0$ from $2\times10^{-6}$~G, the minimum value of  $T_{21}$ profile decreases. For the case when  $B_0=1\times10^{-6}$~G (blue solid line), minimum of $T_{21}$ profile is well below the EDGES upper limit on $T_{21}$ (i.e. $-300$~mK---magenta dashed line). In figure \eqref{plot:1a}, when there is no effect of X-ray heating,  $T_{21}=-300$~mK  corresponds to  $B_0=1.35\times10^{-6}$~G.  Thus, we need to lower $B_0$ values  when the X-ray heating is included to get  desired value of $T_{21}$. As shown in Fig.\eqref{fig:mag_21}, brightness temperature is suppressed  by the increase of the strength of the magnetic field and it can even erase the standard 21 cm signal when the magnetic field strength increases above $\sim1 \times 10^{-6}$~G. This sets the upper limit on the strength of the magnetic field for $m_d=0.1$~GeV and $\hat{\sigma}=6.22\times 10^{-15}~{\rm GeV}^{-2}$.


\section{Conclusion}\label{sec:conc}

Magnetic field in \cite{Minoda:2018gxj} have shown to heat the cosmic gas during the cosmic dawn era by the ambipolar diffusion and the turbulence decay.  Since, it could erase the observed 21~cm absorption signal, one can calculate the upper bound on the magnetic field. One of the promising mechanism to explain the absorption signal of the 21~cm line is to have interaction between the dark-matter and baryons \cite{Barkana:2018lgd,Barkana:2018nd}. In this work, we have shown that in the presence of such an interaction the upper bound on the strength of magnetic fields can significantly be altered. The magnetic-energy converted to the thermal energy, heat both the gas and dark matter. This is an extra heating effect in addition to the drag heating. The drag term heats the DM and baryons, but in the lower range of dark-matter mass ($\ll1$~GeV) it favors cooling of gas compared to heating due to the relative motion between DM and gas. To explain the observed anomaly in the 21 cm signal by the EDGES, a large baryon-DM scattering cross-section is required to balance the magnetic heating effect.  An earlier saturation occurs in baryon-DM cross-section with respect to the DM mass in the presence of the strong magnetic fields.
 Considering upper bound on $\hat{\sigma}-m_d$ by Planck 2015 \cite{ Boddy:2018G} and EDGES upper constraint on $T_{21}$ (-300~mK) at $z=17$ \cite{Bowman:2018yin}, we found upper bound on the magnetic field strength: $B_0=3.48\times10^{-6}$~G, while considering CMB-S4 forecast constraint \cite{Kovetz:2018P} we get $B_0=2.67\times 10^{-6}$~G  for the dark matter mass $\lesssim 10^{-2}$~GeV. \\

{\bf Acknowledgments:} We would like to thank the anonymous reader whose comments has helped us in improving  presentation of our results.


\bibliographystyle{apsrev4-1}

\begin{thebibliography}{68}%
    \makeatletter
    \providecommand \@ifxundefined [1]{%
        \@ifx{#1\undefined}
    }%
    \providecommand \@ifnum [1]{%
        \ifnum #1\expandafter \@firstoftwo
        \else \expandafter \@secondoftwo
        \fi
    }%
    \providecommand \@ifx [1]{%
        \ifx #1\expandafter \@firstoftwo
        \else \expandafter \@secondoftwo
        \fi
    }%
    \providecommand \natexlab [1]{#1}%
    \providecommand \enquote  [1]{``#1''}%
    \providecommand \bibnamefont  [1]{#1}%
    \providecommand \bibfnamefont [1]{#1}%
    \providecommand \citenamefont [1]{#1}%
    \providecommand \href@noop [0]{\@secondoftwo}%
    \providecommand \href [0]{\begingroup \@sanitize@url \@href}%
    \providecommand \@href[1]{\@@startlink{#1}\@@href}%
    \providecommand \@@href[1]{\endgroup#1\@@endlink}%
    \providecommand \@sanitize@url [0]{\catcode `\\12\catcode `\$12\catcode
        `\&12\catcode `\#12\catcode `\^12\catcode `\_12\catcode `\%12\relax}%
    \providecommand \@@startlink[1]{}%
    \providecommand \@@endlink[0]{}%
    \providecommand \url  [0]{\begingroup\@sanitize@url \@url }%
    \providecommand \@url [1]{\endgroup\@href {#1}{\urlprefix }}%
    \providecommand \urlprefix  [0]{URL }%
    \providecommand \Eprint [0]{\href }%
    \providecommand \doibase [0]{http://dx.doi.org/}%
    \providecommand \selectlanguage [0]{\@gobble}%
    \providecommand \bibinfo  [0]{\@secondoftwo}%
    \providecommand \bibfield  [0]{\@secondoftwo}%
    \providecommand \translation [1]{[#1]}%
    \providecommand \BibitemOpen [0]{}%
    \providecommand \bibitemStop [0]{}%
    \providecommand \bibitemNoStop [0]{.\EOS\space}%
    \providecommand \EOS [0]{\spacefactor3000\relax}%
    \providecommand \BibitemShut  [1]{\csname bibitem#1\endcsname}%
    \let\auto@bib@innerbib\@empty
    \bibitem [{\citenamefont {Minoda}\ \emph {et~al.}(2019)\citenamefont {Minoda},
        \citenamefont {Tashiro},\ and\ \citenamefont {Takahashi}}]{Minoda:2018gxj}%
    \BibitemOpen
    \bibfield  {author} {\bibinfo {author} {\bibfnamefont {T.}~\bibnamefont
            {Minoda}}, \bibinfo {author} {\bibfnamefont {H.}~\bibnamefont {Tashiro}}, \
        and\ \bibinfo {author} {\bibfnamefont {T.}~\bibnamefont {Takahashi}},\ }\href
    {\doibase 10.1093/mnras/stz1860} {\bibfield  {journal} {\bibinfo  {journal}
            {MNRAS}\ }\textbf {\bibinfo {volume} {488}},\ \bibinfo {pages} {2001–2005}
        (\bibinfo {year} {2019})}\BibitemShut {NoStop}%
    \bibitem [{\citenamefont {Hogan}\ and\ \citenamefont
        {Rees}(1979)}]{Hogan:1979mj}%
    \BibitemOpen
    \bibfield  {author} {\bibinfo {author} {\bibfnamefont {C.~J.}\ \bibnamefont
            {Hogan}}\ and\ \bibinfo {author} {\bibfnamefont {M.~J.}\ \bibnamefont
            {Rees}},\ }\href {\doibase 10.1093/mnras/188.4.791} {\bibfield  {journal}
        {\bibinfo  {journal} {MNRAS}\ }\textbf {\bibinfo {volume} {188}},\ \bibinfo
        {pages} {791} (\bibinfo {year} {1979})}\BibitemShut {NoStop}%
    \bibitem [{\citenamefont {Scott}\ and\ \citenamefont
        {Rees}(1990)}]{Scott:1990mj}%
    \BibitemOpen
    \bibfield  {author} {\bibinfo {author} {\bibfnamefont {D.}~\bibnamefont
            {Scott}}\ and\ \bibinfo {author} {\bibfnamefont {M.~J.}\ \bibnamefont
            {Rees}},\ }\href@noop {} {\bibfield  {journal} {\bibinfo  {journal} {MNRAS}\
        }\textbf {\bibinfo {volume} {247}},\ \bibinfo {pages} {510} (\bibinfo {year}
        {1990})}\BibitemShut {NoStop}%
    \bibitem [{\citenamefont {Fialkov}\ \emph {et~al.}(2014)\citenamefont
        {Fialkov}, \citenamefont {Barkana},\ and\ \citenamefont
        {Visbal}}]{Fialkov:2014kta}%
    \BibitemOpen
    \bibfield  {author} {\bibinfo {author} {\bibfnamefont {A.}~\bibnamefont
            {Fialkov}}, \bibinfo {author} {\bibfnamefont {R.}~\bibnamefont {Barkana}}, \
        and\ \bibinfo {author} {\bibfnamefont {E.}~\bibnamefont {Visbal}},\ }\href
    {\doibase 10.1038/nature12999} {\bibfield  {journal} {\bibinfo  {journal}
            {Nature}\ }\textbf {\bibinfo {volume} {506}},\ \bibinfo {pages} {197}
        (\bibinfo {year} {2014})},\ \Eprint {http://arxiv.org/abs/1402.0940}
    {arXiv:1402.0940 [astro-ph.CO]} \BibitemShut {NoStop}%
    \bibitem [{\citenamefont {Bowman}\ \emph {et~al.}(2018)\citenamefont {Bowman},
        \citenamefont {Rogers}, \citenamefont {Monsalve}, \citenamefont {Mozdzen},\
        and\ \citenamefont {Mahesh}}]{Bowman:2018yin}%
    \BibitemOpen
    \bibfield  {author} {\bibinfo {author} {\bibfnamefont {J.~D.}\ \bibnamefont
            {Bowman}}, \bibinfo {author} {\bibfnamefont {A.~E.~E.}\ \bibnamefont
            {Rogers}}, \bibinfo {author} {\bibfnamefont {R.~A.}\ \bibnamefont
            {Monsalve}}, \bibinfo {author} {\bibfnamefont {T.~J.}\ \bibnamefont
            {Mozdzen}}, \ and\ \bibinfo {author} {\bibfnamefont {N.}~\bibnamefont
            {Mahesh}},\ }\href {\doibase 10.1038/nature25792} {\bibfield  {journal}
        {\bibinfo  {journal} {Nature}\ }\textbf {\bibinfo {volume} {555}},\ \bibinfo
        {pages} {67} (\bibinfo {year} {2018})},\ \Eprint
    {http://arxiv.org/abs/1810.05912} {arXiv:1810.05912 [astro-ph.CO]}
    \BibitemShut {NoStop}%
    \bibitem [{\citenamefont {Moroi}\ \emph {et~al.}(2018)\citenamefont {Moroi},
        \citenamefont {Nakayama},\ and\ \citenamefont {Tang}}]{Moroi:2018vci}%
    \BibitemOpen
    \bibfield  {author} {\bibinfo {author} {\bibfnamefont {T.}~\bibnamefont
            {Moroi}}, \bibinfo {author} {\bibfnamefont {K.}~\bibnamefont {Nakayama}}, \
        and\ \bibinfo {author} {\bibfnamefont {Y.}~\bibnamefont {Tang}},\ }\href
    {\doibase 10.1016/j.physletb.2018.07.002} {\bibfield  {journal} {\bibinfo
            {journal} {Phys. Lett.}\ }\textbf {\bibinfo {volume} {B783}},\ \bibinfo
        {pages} {301} (\bibinfo {year} {2018})},\ \Eprint
    {http://arxiv.org/abs/1804.10378} {arXiv:1804.10378 [hep-ph]} \BibitemShut
    {NoStop}%
    \bibitem [{\citenamefont {Fraser}\ \emph {et~al.}(2018)\citenamefont {Fraser},
        \citenamefont {Hektor}, \citenamefont {Hütsi}, \citenamefont {Kannike},
        \citenamefont {Marzo}, \citenamefont {Marzola}, \citenamefont {Racioppi},
        \citenamefont {Raidal}, \citenamefont {Spethmann}, \citenamefont {Vaskonen},\
        and\ \citenamefont {Veermäe}}]{FRASER2018159}%
    \BibitemOpen
    \bibfield  {author} {\bibinfo {author} {\bibfnamefont {S.}~\bibnamefont
            {Fraser}}, \bibinfo {author} {\bibfnamefont {A.}~\bibnamefont {Hektor}},
        \bibinfo {author} {\bibfnamefont {G.}~\bibnamefont {Hütsi}}, \bibinfo
        {author} {\bibfnamefont {K.}~\bibnamefont {Kannike}}, \bibinfo {author}
        {\bibfnamefont {C.}~\bibnamefont {Marzo}}, \bibinfo {author} {\bibfnamefont
            {L.}~\bibnamefont {Marzola}}, \bibinfo {author} {\bibfnamefont
            {A.}~\bibnamefont {Racioppi}}, \bibinfo {author} {\bibfnamefont
            {M.}~\bibnamefont {Raidal}}, \bibinfo {author} {\bibfnamefont
            {C.}~\bibnamefont {Spethmann}}, \bibinfo {author} {\bibfnamefont
            {V.}~\bibnamefont {Vaskonen}}, \ and\ \bibinfo {author} {\bibfnamefont
            {H.}~\bibnamefont {Veermäe}},\ }\href {\doibase
        https://doi.org/10.1016/j.physletb.2018.08.035} {\bibfield  {journal}
        {\bibinfo  {journal} {Physics Letters B}\ }\textbf {\bibinfo {volume}
            {785}},\ \bibinfo {pages} {159 } (\bibinfo {year} {2018})}\BibitemShut
    {NoStop}%
    \bibitem [{\citenamefont {Pospelov}\ \emph {et~al.}(2018)\citenamefont
        {Pospelov}, \citenamefont {Pradler}, \citenamefont {Ruderman},\ and\
        \citenamefont {Urbano}}]{PhysRevLett.121.031103}%
    \BibitemOpen
    \bibfield  {author} {\bibinfo {author} {\bibfnamefont {M.}~\bibnamefont
            {Pospelov}}, \bibinfo {author} {\bibfnamefont {J.}~\bibnamefont {Pradler}},
        \bibinfo {author} {\bibfnamefont {J.~T.}\ \bibnamefont {Ruderman}}, \ and\
        \bibinfo {author} {\bibfnamefont {A.}~\bibnamefont {Urbano}},\ }\href
    {\doibase 10.1103/PhysRevLett.121.031103} {\bibfield  {journal} {\bibinfo
            {journal} {Phys. Rev. Lett.}\ }\textbf {\bibinfo {volume} {121}},\ \bibinfo
        {pages} {031103} (\bibinfo {year} {2018})}\BibitemShut {NoStop}%
    \bibitem [{\citenamefont {Liu}\ \emph {et~al.}(2019)\citenamefont {Liu},
        \citenamefont {Outmezguine}, \citenamefont {Redigolo},\ and\ \citenamefont
        {Volansky}}]{Liu:2019H}%
    \BibitemOpen
    \bibfield  {author} {\bibinfo {author} {\bibfnamefont {H.}~\bibnamefont
            {Liu}}, \bibinfo {author} {\bibfnamefont {N.~J.}\ \bibnamefont
            {Outmezguine}}, \bibinfo {author} {\bibfnamefont {D.}~\bibnamefont
            {Redigolo}}, \ and\ \bibinfo {author} {\bibfnamefont {T.}~\bibnamefont
            {Volansky}},\ }\href {\doibase 10.1103/PhysRevD.100.123011} {\bibfield
        {journal} {\bibinfo  {journal} {Phys. Rev. D}\ }\textbf {\bibinfo {volume}
            {100}},\ \bibinfo {pages} {123011} (\bibinfo {year} {2019})}\BibitemShut
    {NoStop}%
    \bibitem [{\citenamefont {Chuzhoy}\ and\ \citenamefont
        {Shapiro}(2007)}]{Chuzhoy_2007}%
    \BibitemOpen
    \bibfield  {author} {\bibinfo {author} {\bibfnamefont {L.}~\bibnamefont
            {Chuzhoy}}\ and\ \bibinfo {author} {\bibfnamefont {P.~R.}\ \bibnamefont
            {Shapiro}},\ }\href {\doibase 10.1086/510146} {\bibfield  {journal} {\bibinfo
            {journal} {The Astrophysical Journal}\ }\textbf {\bibinfo {volume} {655}},\
        \bibinfo {pages} {843} (\bibinfo {year} {2007})}\BibitemShut {NoStop}%
    \bibitem [{\citenamefont {Chuzhoy}\ and\ \citenamefont
        {Shapiro}(2006)}]{Chuzhoy2006}%
    \BibitemOpen
    \bibfield  {author} {\bibinfo {author} {\bibfnamefont {L.}~\bibnamefont
            {Chuzhoy}}\ and\ \bibinfo {author} {\bibfnamefont {P.~R.}\ \bibnamefont
            {Shapiro}},\ }\href {\doibase 10.1086/507670} {\bibfield  {journal} {\bibinfo
            {journal} {The Astrophysical Journal}\ }\textbf {\bibinfo {volume} {651}},\
        \bibinfo {pages} {1} (\bibinfo {year} {2006})}\BibitemShut {NoStop}%
    \bibitem [{\citenamefont {Barkana}\ \emph {et~al.}(2018)\citenamefont
        {Barkana}, \citenamefont {Outmezguine}, \citenamefont {Redigolo},\ and\
        \citenamefont {Volansky}}]{Barkana:2018nd}%
    \BibitemOpen
    \bibfield  {author} {\bibinfo {author} {\bibfnamefont {R.}~\bibnamefont
            {Barkana}}, \bibinfo {author} {\bibfnamefont {N.~J.}\ \bibnamefont
            {Outmezguine}}, \bibinfo {author} {\bibfnamefont {D.}~\bibnamefont
            {Redigolo}}, \ and\ \bibinfo {author} {\bibfnamefont {T.}~\bibnamefont
            {Volansky}},\ }\href {\doibase 10.1103/PhysRevD.98.103005} {\bibfield
        {journal} {\bibinfo  {journal} {Phys. Rev. D}\ }\textbf {\bibinfo {volume}
            {98}},\ \bibinfo {pages} {103005} (\bibinfo {year} {2018})}\BibitemShut
    {NoStop}%
    \bibitem [{\citenamefont {Barkana}(2018)}]{Barkana:2018lgd}%
    \BibitemOpen
    \bibfield  {author} {\bibinfo {author} {\bibfnamefont {R.}~\bibnamefont
            {Barkana}},\ }\href {\doibase 10.1038/nature25791} {\bibfield  {journal}
        {\bibinfo  {journal} {Nature}\ }\textbf {\bibinfo {volume} {555}},\ \bibinfo
        {pages} {71} (\bibinfo {year} {2018})},\ \Eprint
    {http://arxiv.org/abs/1803.06698} {arXiv:1803.06698 [astro-ph.CO]}
    \BibitemShut {NoStop}%
    \bibitem [{\citenamefont {Tashiro}\ \emph {et~al.}(2014)\citenamefont
        {Tashiro}, \citenamefont {Kadota},\ and\ \citenamefont
        {Silk}}]{Tashiro:2014tsa}%
    \BibitemOpen
    \bibfield  {author} {\bibinfo {author} {\bibfnamefont {H.}~\bibnamefont
            {Tashiro}}, \bibinfo {author} {\bibfnamefont {K.}~\bibnamefont {Kadota}}, \
        and\ \bibinfo {author} {\bibfnamefont {J.}~\bibnamefont {Silk}},\ }\href
    {\doibase 10.1103/PhysRevD.90.083522} {\bibfield  {journal} {\bibinfo
            {journal} {Phys. Rev.}\ }\textbf {\bibinfo {volume} {D90}},\ \bibinfo {pages}
        {083522} (\bibinfo {year} {2014})},\ \Eprint {http://arxiv.org/abs/1408.2571}
    {arXiv:1408.2571 [astro-ph.CO]} \BibitemShut {NoStop}%
    \bibitem [{\citenamefont {Dvorkin}\ \emph {et~al.}(2014)\citenamefont
        {Dvorkin}, \citenamefont {Blum},\ and\ \citenamefont
        {Kamionkowski}}]{Dvorkin:2013cea}%
    \BibitemOpen
    \bibfield  {author} {\bibinfo {author} {\bibfnamefont {C.}~\bibnamefont
            {Dvorkin}}, \bibinfo {author} {\bibfnamefont {K.}~\bibnamefont {Blum}}, \
        and\ \bibinfo {author} {\bibfnamefont {M.}~\bibnamefont {Kamionkowski}},\
    }\href {\doibase 10.1103/PhysRevD.89.023519} {\bibfield  {journal} {\bibinfo
            {journal} {Phys. Rev.}\ }\textbf {\bibinfo {volume} {D89}},\ \bibinfo {pages}
        {023519} (\bibinfo {year} {2014})},\ \Eprint {http://arxiv.org/abs/1311.2937}
    {arXiv:1311.2937 [astro-ph.CO]} \BibitemShut {NoStop}%
    \bibitem [{\citenamefont {Berlin}\ \emph {et~al.}(2018)\citenamefont {Berlin},
        \citenamefont {Hooper}, \citenamefont {Krnjaic},\ and\ \citenamefont
        {McDermott}}]{Berlin:2018sjs}%
    \BibitemOpen
    \bibfield  {author} {\bibinfo {author} {\bibfnamefont {A.}~\bibnamefont
            {Berlin}}, \bibinfo {author} {\bibfnamefont {D.}~\bibnamefont {Hooper}},
        \bibinfo {author} {\bibfnamefont {G.}~\bibnamefont {Krnjaic}}, \ and\
        \bibinfo {author} {\bibfnamefont {S.~D.}\ \bibnamefont {McDermott}},\ }\href
    {\doibase 10.1103/PhysRevLett.121.011102} {\bibfield  {journal} {\bibinfo
            {journal} {Phys. Rev. Lett.}\ }\textbf {\bibinfo {volume} {121}},\ \bibinfo
        {pages} {011102} (\bibinfo {year} {2018})},\ \Eprint
    {http://arxiv.org/abs/1803.02804} {arXiv:1803.02804 [hep-ph]} \BibitemShut
    {NoStop}%
    \bibitem [{\citenamefont {Creque-Sarbinowski}\ \emph
        {et~al.}(2019)\citenamefont {Creque-Sarbinowski}, \citenamefont {Ji},
        \citenamefont {Kovetz},\ and\ \citenamefont
        {Kamionkowski}}]{Creque-Sarbinowski:2019mcm}%
    \BibitemOpen
    \bibfield  {author} {\bibinfo {author} {\bibfnamefont {C.}~\bibnamefont
            {Creque-Sarbinowski}}, \bibinfo {author} {\bibfnamefont {L.}~\bibnamefont
            {Ji}}, \bibinfo {author} {\bibfnamefont {E.~D.}\ \bibnamefont {Kovetz}}, \
        and\ \bibinfo {author} {\bibfnamefont {M.}~\bibnamefont {Kamionkowski}},\
    }\href@noop {} {\  (\bibinfo {year} {2019})},\ \Eprint
    {http://arxiv.org/abs/1903.09154} {arXiv:1903.09154 [astro-ph.CO]}
    \BibitemShut {NoStop}%
    \bibitem [{\citenamefont {Kronberg}(1994)}]{Kronberg:1994pp}%
    \BibitemOpen
    \bibfield  {author} {\bibinfo {author} {\bibfnamefont {P.}~\bibnamefont
            {Kronberg}},\ }\href {\doibase 10.1088/0034-4885/57/4/001} {\bibfield
        {journal} {\bibinfo  {journal} {Reports on Progress in Physics}\ }\textbf
        {\bibinfo {volume} {57}},\ \bibinfo {pages} {325} (\bibinfo {year}
        {1994})}\BibitemShut {NoStop}%
    \bibitem [{\citenamefont {Neronov}\ and\ \citenamefont
        {Vovk}(2010)}]{Neronov:1900zz}%
    \BibitemOpen
    \bibfield  {author} {\bibinfo {author} {\bibfnamefont {A.}~\bibnamefont
            {Neronov}}\ and\ \bibinfo {author} {\bibfnamefont {I.}~\bibnamefont {Vovk}},\
    }\href {\doibase 10.1126/science.1184192} {\bibfield  {journal} {\bibinfo
            {journal} {Science}\ }\textbf {\bibinfo {volume} {328}},\ \bibinfo {pages}
        {73} (\bibinfo {year} {2010})},\ \Eprint {http://arxiv.org/abs/1006.3504}
    {arXiv:1006.3504 [astro-ph.HE]} \BibitemShut {NoStop}%
    \bibitem [{\citenamefont {Trivedi}\ \emph {et~al.}(2012)\citenamefont
        {Trivedi}, \citenamefont {Seshadri},\ and\ \citenamefont
        {Subramanian}}]{Trivedi:2012ssp}%
    \BibitemOpen
    \bibfield  {author} {\bibinfo {author} {\bibfnamefont {P.}~\bibnamefont
            {Trivedi}}, \bibinfo {author} {\bibfnamefont {T.~R.}\ \bibnamefont
            {Seshadri}}, \ and\ \bibinfo {author} {\bibfnamefont {K.}~\bibnamefont
            {Subramanian}},\ }\href {\doibase 10.1103/PhysRevLett.108.231301} {\bibfield
        {journal} {\bibinfo  {journal} {Phys. Rev. Lett.}\ }\textbf {\bibinfo
            {volume} {108}},\ \bibinfo {pages} {231301} (\bibinfo {year}
        {2012})}\BibitemShut {NoStop}%
    \bibitem [{\citenamefont {Trivedi}\ \emph {et~al.}(2014)\citenamefont
        {Trivedi}, \citenamefont {Subramanian},\ and\ \citenamefont
        {Seshadri}}]{Trivedi:2013wqa}%
    \BibitemOpen
    \bibfield  {author} {\bibinfo {author} {\bibfnamefont {P.}~\bibnamefont
            {Trivedi}}, \bibinfo {author} {\bibfnamefont {K.}~\bibnamefont
            {Subramanian}}, \ and\ \bibinfo {author} {\bibfnamefont {T.~R.}\ \bibnamefont
            {Seshadri}},\ }\href {\doibase 10.1103/PhysRevD.89.043523} {\bibfield
        {journal} {\bibinfo  {journal} {Phys. Rev.}\ }\textbf {\bibinfo {volume}
            {D89}},\ \bibinfo {pages} {043523} (\bibinfo {year} {2014})},\ \Eprint
    {http://arxiv.org/abs/1312.5308} {arXiv:1312.5308 [astro-ph.CO]} \BibitemShut
    {NoStop}%
    \bibitem [{\citenamefont {Sethi}\ and\ \citenamefont
        {Subramanian}(2005)}]{Sethi:2004pe}%
    \BibitemOpen
    \bibfield  {author} {\bibinfo {author} {\bibfnamefont {S.~K.}\ \bibnamefont
            {Sethi}}\ and\ \bibinfo {author} {\bibfnamefont {K.}~\bibnamefont
            {Subramanian}},\ }\href {\doibase 10.1111/j.1365-2966.2004.08520.x}
    {\bibfield  {journal} {\bibinfo  {journal} {Mon. Not. Roy. Astron. Soc.}\
        }\textbf {\bibinfo {volume} {356}},\ \bibinfo {pages} {778} (\bibinfo {year}
        {2005})},\ \Eprint {http://arxiv.org/abs/astro-ph/0405413}
    {arXiv:astro-ph/0405413 [astro-ph]} \BibitemShut {NoStop}%
    \bibitem [{\citenamefont {Ade}\ \emph {et~al.}(2016)\citenamefont {Ade} \emph
        {et~al.}}]{Ade:2015cva}%
    \BibitemOpen
    \bibfield  {author} {\bibinfo {author} {\bibfnamefont {P.~A.~R.}\
            \bibnamefont {Ade}} \emph {et~al.} (\bibinfo {collaboration} {Planck}),\
    }\href {\doibase 10.1051/0004-6361/201525821} {\bibfield  {journal} {\bibinfo
            {journal} {Astron. Astrophys.}\ }\textbf {\bibinfo {volume} {594}},\
        \bibinfo {pages} {A19} (\bibinfo {year} {2016})},\ \Eprint
    {http://arxiv.org/abs/1502.01594} {arXiv:1502.01594 [astro-ph.CO]}
    \BibitemShut {NoStop}%
    \bibitem [{\citenamefont {Cheng}\ \emph {et~al.}(1996)\citenamefont {Cheng},
        \citenamefont {Olinto}, \citenamefont {Schramm},\ and\ \citenamefont
        {Truran}}]{Cheng:1996vn}%
    \BibitemOpen
    \bibfield  {author} {\bibinfo {author} {\bibfnamefont {B.}~\bibnamefont
            {Cheng}}, \bibinfo {author} {\bibfnamefont {A.~V.}\ \bibnamefont {Olinto}},
        \bibinfo {author} {\bibfnamefont {D.~N.}\ \bibnamefont {Schramm}}, \ and\
        \bibinfo {author} {\bibfnamefont {J.~W.}\ \bibnamefont {Truran}},\ }\href
    {\doibase 10.1103/PhysRevD.54.4714} {\bibfield  {journal} {\bibinfo
            {journal} {Phys. Rev. D}\ }\textbf {\bibinfo {volume} {54}},\ \bibinfo
        {pages} {4714} (\bibinfo {year} {1996})}\BibitemShut {NoStop}%
    \bibitem [{\citenamefont {Grasso}\ and\ \citenamefont
        {Rubinstein}(2001)}]{Grasso:2000wj}%
    \BibitemOpen
    \bibfield  {author} {\bibinfo {author} {\bibfnamefont {D.}~\bibnamefont
            {Grasso}}\ and\ \bibinfo {author} {\bibfnamefont {H.~R.}\ \bibnamefont
            {Rubinstein}},\ }\href {\doibase 10.1016/S0370-1573(00)00110-1} {\bibfield
        {journal} {\bibinfo  {journal} {Phys. Rept.}\ }\textbf {\bibinfo {volume}
            {348}},\ \bibinfo {pages} {163} (\bibinfo {year} {2001})},\ \Eprint
    {http://arxiv.org/abs/astro-ph/0009061} {arXiv:astro-ph/0009061 [astro-ph]}
    \BibitemShut {NoStop}%
    \bibitem [{\citenamefont {Matese}\ and\ \citenamefont
        {O'Connell}(1969)}]{Matese:1969cj}%
    \BibitemOpen
    \bibfield  {author} {\bibinfo {author} {\bibfnamefont {J.~J.}\ \bibnamefont
            {Matese}}\ and\ \bibinfo {author} {\bibfnamefont {R.~F.}\ \bibnamefont
            {O'Connell}},\ }\href {\doibase 10.1103/PhysRev.180.1289} {\bibfield
        {journal} {\bibinfo  {journal} {Phys. Rev.}\ }\textbf {\bibinfo {volume}
            {180}},\ \bibinfo {pages} {1289} (\bibinfo {year} {1969})}\BibitemShut
    {NoStop}%
    \bibitem [{\citenamefont {GREENSTEIN}(1969)}]{Greenstein:1969}%
    \BibitemOpen
    \bibfield  {author} {\bibinfo {author} {\bibfnamefont {G.}~\bibnamefont
            {GREENSTEIN}},\ }\href {\doibase 10.1038/223938b0} {\bibfield  {journal}
        {\bibinfo  {journal} {Nature}\ }\textbf {\bibinfo {volume} {223}},\ \bibinfo
        {pages} {938} (\bibinfo {year} {1969})}\BibitemShut {NoStop}%
    \bibitem [{\citenamefont {Tashiro}\ and\ \citenamefont
        {Sugiyama}(2006{\natexlab{a}})}]{Tashiro:2005ua}%
    \BibitemOpen
    \bibfield  {author} {\bibinfo {author} {\bibfnamefont {H.}~\bibnamefont
            {Tashiro}}\ and\ \bibinfo {author} {\bibfnamefont {N.}~\bibnamefont
            {Sugiyama}},\ }\href {\doibase 10.1111/j.1365-2966.2006.10178.x} {\bibfield
        {journal} {\bibinfo  {journal} {Mon. Not. Roy. Astron. Soc.}\ }\textbf
        {\bibinfo {volume} {368}},\ \bibinfo {pages} {965} (\bibinfo {year}
        {2006}{\natexlab{a}})},\ \Eprint {http://arxiv.org/abs/astro-ph/0512626}
    {arXiv:astro-ph/0512626 [astro-ph]} \BibitemShut {NoStop}%
    \bibitem [{\citenamefont {Schleicher}\ \emph {et~al.}(2008)\citenamefont
        {Schleicher}, \citenamefont {Banerjee},\ and\ \citenamefont
        {Klesser}}]{Schleicher:2008aa}%
    \BibitemOpen
    \bibfield  {author} {\bibinfo {author} {\bibfnamefont {D.~R.~G.}\
            \bibnamefont {Schleicher}}, \bibinfo {author} {\bibfnamefont
            {R.}~\bibnamefont {Banerjee}}, \ and\ \bibinfo {author} {\bibfnamefont
            {R.~S.}\ \bibnamefont {Klesser}},\ }\href {\doibase
        10.1103/PhysRevD.78.083005} {\bibfield  {journal} {\bibinfo  {journal} {Phys.
                Rev.}\ }\textbf {\bibinfo {volume} {D78}},\ \bibinfo {pages} {083005}
        (\bibinfo {year} {2008})},\ \Eprint {http://arxiv.org/abs/0807.3802}
    {arXiv:0807.3802 [astro-ph]} \BibitemShut {NoStop}%
    \bibitem [{\citenamefont {Chluba}\ \emph {et~al.}(2015)\citenamefont {Chluba},
        \citenamefont {Paoletti}, \citenamefont {Finelli},\ and\ \citenamefont
        {Rubi{\~{n}}o-Mart{\'{i}}n}}]{Chluba2015}%
    \BibitemOpen
    \bibfield  {author} {\bibinfo {author} {\bibfnamefont {J.}~\bibnamefont
            {Chluba}}, \bibinfo {author} {\bibfnamefont {D.}~\bibnamefont {Paoletti}},
        \bibinfo {author} {\bibfnamefont {F.}~\bibnamefont {Finelli}}, \ and\
        \bibinfo {author} {\bibfnamefont {J.~A.}\ \bibnamefont
            {Rubi{\~{n}}o-Mart{\'{i}}n}},\ }\href {\doibase 10.1093/mnras/stv1096}
    {\bibfield  {journal} {\bibinfo  {journal} {Monthly Notices of the Royal
                Astronomical Society}\ }\textbf {\bibinfo {volume} {451}},\ \bibinfo {pages}
        {2244} (\bibinfo {year} {2015})}\BibitemShut {NoStop}%
    \bibitem [{\citenamefont {Sethi}\ \emph {et~al.}(2008)\citenamefont {Sethi},
        \citenamefont {Nath},\ and\ \citenamefont {Subramanian}}]{Sethi:2008eq}%
    \BibitemOpen
    \bibfield  {author} {\bibinfo {author} {\bibfnamefont {S.~K.}\ \bibnamefont
            {Sethi}}, \bibinfo {author} {\bibfnamefont {B.~B.}\ \bibnamefont {Nath}}, \
        and\ \bibinfo {author} {\bibfnamefont {K.}~\bibnamefont {Subramanian}},\
    }\href {\doibase 10.1111/j.1365-2966.2008.13302.x} {\bibfield  {journal}
        {\bibinfo  {journal} {Mon. Not. Roy. Astron. Soc.}\ }\textbf {\bibinfo
            {volume} {387}},\ \bibinfo {pages} {1589} (\bibinfo {year} {2008})},\ \Eprint
    {http://arxiv.org/abs/0804.3473} {arXiv:0804.3473 [astro-ph]} \BibitemShut
    {NoStop}%
    \bibitem [{\citenamefont {Shu}(1992)}]{Shu:1992fh}%
    \BibitemOpen
    \bibfield  {author} {\bibinfo {author} {\bibfnamefont {F.~H.}\ \bibnamefont
            {Shu}},\ }\href@noop {} {\emph {\bibinfo {title} {The physics of
                astrophysics.~Volume II: Gas dynamics.,~ University Science Books, Mill
                Valley, CA (USA), ISBN 0-935702-65-2}}}\ (\bibinfo {year} {1992})\BibitemShut
    {NoStop}%
    \bibitem [{\citenamefont {Oklop{\v{c}}i{\'{c}}}\ and\ \citenamefont
        {Hirata}(2013)}]{Oklop_i__2013}%
    \BibitemOpen
    \bibfield  {author} {\bibinfo {author} {\bibfnamefont {A.}~\bibnamefont
            {Oklop{\v{c}}i{\'{c}}}}\ and\ \bibinfo {author} {\bibfnamefont {C.~M.}\
            \bibnamefont {Hirata}},\ }\href {\doibase 10.1088/0004-637x/779/2/146}
    {\bibfield  {journal} {\bibinfo  {journal} {The Astrophysical Journal}\
        }\textbf {\bibinfo {volume} {779}},\ \bibinfo {pages} {146} (\bibinfo {year}
        {2013})}\BibitemShut {NoStop}%
    \bibitem [{\citenamefont {Chuzhoy}\ \emph {et~al.}(2007)\citenamefont
        {Chuzhoy}, \citenamefont {Kuhlen},\ and\ \citenamefont
        {Shapiro}}]{Chuzhoy2007}%
    \BibitemOpen
    \bibfield  {author} {\bibinfo {author} {\bibfnamefont {L.}~\bibnamefont
            {Chuzhoy}}, \bibinfo {author} {\bibfnamefont {M.}~\bibnamefont {Kuhlen}}, \
        and\ \bibinfo {author} {\bibfnamefont {P.~R.}\ \bibnamefont {Shapiro}},\
    }\href {\doibase 10.1086/521438} {\bibfield  {journal} {\bibinfo  {journal}
            {The Astrophysical Journal}\ }\textbf {\bibinfo {volume} {665}},\ \bibinfo
        {pages} {L85} (\bibinfo {year} {2007})}\BibitemShut {NoStop}%
    \bibitem [{\citenamefont {Ghara}\ and\ \citenamefont
        {Mellema}(2019)}]{Ghara2019}%
    \BibitemOpen
    \bibfield  {author} {\bibinfo {author} {\bibfnamefont {R.}~\bibnamefont
            {Ghara}}\ and\ \bibinfo {author} {\bibfnamefont {G.}~\bibnamefont
            {Mellema}},\ }\href {\doibase 10.1093/mnras/stz3513} {\bibfield  {journal}
        {\bibinfo  {journal} {Monthly Notices of the Royal Astronomical Society}\
        }\textbf {\bibinfo {volume} {492}},\ \bibinfo {pages} {634} (\bibinfo {year}
        {2019})}\BibitemShut {NoStop}%
    \bibitem [{\citenamefont {Slatyer}(2016{\natexlab{a}})}]{PhysRevD.93.023527}%
    \BibitemOpen
    \bibfield  {author} {\bibinfo {author} {\bibfnamefont {T.~R.}\ \bibnamefont
            {Slatyer}},\ }\href {\doibase 10.1103/PhysRevD.93.023527} {\bibfield
        {journal} {\bibinfo  {journal} {Phys. Rev. D}\ }\textbf {\bibinfo {volume}
            {93}},\ \bibinfo {pages} {023527} (\bibinfo {year}
        {2016}{\natexlab{a}})}\BibitemShut {NoStop}%
    \bibitem [{\citenamefont {Slatyer}(2016{\natexlab{b}})}]{PhysRevD.93.023521}%
    \BibitemOpen
    \bibfield  {author} {\bibinfo {author} {\bibfnamefont {T.~R.}\ \bibnamefont
            {Slatyer}},\ }\href {\doibase 10.1103/PhysRevD.93.023521} {\bibfield
        {journal} {\bibinfo  {journal} {Phys. Rev. D}\ }\textbf {\bibinfo {volume}
            {93}},\ \bibinfo {pages} {023521} (\bibinfo {year}
        {2016}{\natexlab{b}})}\BibitemShut {NoStop}%
    \bibitem [{\citenamefont {D'Amico}\ \emph {et~al.}(2018)\citenamefont
        {D'Amico}, \citenamefont {Panci},\ and\ \citenamefont
        {Strumia}}]{PhysRevLett.121.011103}%
    \BibitemOpen
    \bibfield  {author} {\bibinfo {author} {\bibfnamefont {G.}~\bibnamefont
            {D'Amico}}, \bibinfo {author} {\bibfnamefont {P.}~\bibnamefont {Panci}}, \
        and\ \bibinfo {author} {\bibfnamefont {A.}~\bibnamefont {Strumia}},\ }\href
    {\doibase 10.1103/PhysRevLett.121.011103} {\bibfield  {journal} {\bibinfo
            {journal} {Phys. Rev. Lett.}\ }\textbf {\bibinfo {volume} {121}},\ \bibinfo
        {pages} {011103} (\bibinfo {year} {2018})}\BibitemShut {NoStop}%
    \bibitem [{\citenamefont {Liu}\ and\ \citenamefont
        {Slatyer}(2018)}]{Liu:2018uzy}%
    \BibitemOpen
    \bibfield  {author} {\bibinfo {author} {\bibfnamefont {H.}~\bibnamefont
            {Liu}}\ and\ \bibinfo {author} {\bibfnamefont {T.~R.}\ \bibnamefont
            {Slatyer}},\ }\href {\doibase 10.1103/PhysRevD.98.023501} {\bibfield
        {journal} {\bibinfo  {journal} {Phys. Rev.}\ }\textbf {\bibinfo {volume}
            {D98}},\ \bibinfo {pages} {023501} (\bibinfo {year} {2018})},\ \Eprint
    {http://arxiv.org/abs/1803.09739} {arXiv:1803.09739 [astro-ph.CO]}
    \BibitemShut {NoStop}%
    \bibitem [{\citenamefont {Mitridate}\ and\ \citenamefont
        {Podo}(2018)}]{Mitridate_2018}%
    \BibitemOpen
    \bibfield  {author} {\bibinfo {author} {\bibfnamefont {A.}~\bibnamefont
            {Mitridate}}\ and\ \bibinfo {author} {\bibfnamefont {A.}~\bibnamefont
            {Podo}},\ }\href {\doibase 10.1088/1475-7516/2018/05/069} {\bibfield
        {journal} {\bibinfo  {journal} {Journal of Cosmology and Astroparticle
                Physics}\ }\textbf {\bibinfo {volume} {2018}},\ \bibinfo {pages} {069}
        (\bibinfo {year} {2018})}\BibitemShut {NoStop}%
    \bibitem [{\citenamefont {Slatyer}\ and\ \citenamefont
        {Wu}(2017)}]{PhysRevD.95.023010}%
    \BibitemOpen
    \bibfield  {author} {\bibinfo {author} {\bibfnamefont {T.~R.}\ \bibnamefont
            {Slatyer}}\ and\ \bibinfo {author} {\bibfnamefont {C.-L.}\ \bibnamefont
            {Wu}},\ }\href {\doibase 10.1103/PhysRevD.95.023010} {\bibfield  {journal}
        {\bibinfo  {journal} {Phys. Rev. D}\ }\textbf {\bibinfo {volume} {95}},\
        \bibinfo {pages} {023010} (\bibinfo {year} {2017})}\BibitemShut {NoStop}%
    \bibitem [{\citenamefont {Bhatt}\ \emph {et~al.}(2019)\citenamefont {Bhatt},
        \citenamefont {Mishra},\ and\ \citenamefont {Nayak}}]{Bhatt:2019qbq}%
    \BibitemOpen
    \bibfield  {author} {\bibinfo {author} {\bibfnamefont {J.~R.}\ \bibnamefont
            {Bhatt}}, \bibinfo {author} {\bibfnamefont {A.~K.}\ \bibnamefont {Mishra}}, \
        and\ \bibinfo {author} {\bibfnamefont {A.~C.}\ \bibnamefont {Nayak}},\
    }\href@noop {} {\  (\bibinfo {year} {2019})},\ \Eprint
    {http://arxiv.org/abs/1901.08451} {arXiv:1901.08451 [astro-ph.CO]}
    \BibitemShut {NoStop}%
    \bibitem [{\citenamefont {Turner}\ and\ \citenamefont
        {Widrow}(1988)}]{turner:1988mw}%
    \BibitemOpen
    \bibfield  {author} {\bibinfo {author} {\bibfnamefont {M.~S.}\ \bibnamefont
            {Turner}}\ and\ \bibinfo {author} {\bibfnamefont {L.~M.}\ \bibnamefont
            {Widrow}},\ }\href {\doibase 10.1103/PhysRevD.37.2743} {\bibfield  {journal}
        {\bibinfo  {journal} {Phys. Rev. D}\ }\textbf {\bibinfo {volume} {37}},\
        \bibinfo {pages} {2743} (\bibinfo {year} {1988})}\BibitemShut {NoStop}%
    \bibitem [{\citenamefont {Sharma}\ \emph {et~al.}(2018)\citenamefont {Sharma},
        \citenamefont {Subramanian},\ and\ \citenamefont
        {Seshadri}}]{Sharma:2018kgs}%
    \BibitemOpen
    \bibfield  {author} {\bibinfo {author} {\bibfnamefont {R.}~\bibnamefont
            {Sharma}}, \bibinfo {author} {\bibfnamefont {K.}~\bibnamefont {Subramanian}},
        \ and\ \bibinfo {author} {\bibfnamefont {T.~R.}\ \bibnamefont {Seshadri}},\
    }\href {\doibase 10.1103/PhysRevD.97.083503} {\bibfield  {journal} {\bibinfo
            {journal} {Phys. Rev.}\ }\textbf {\bibinfo {volume} {D97}},\ \bibinfo {pages}
        {083503} (\bibinfo {year} {2018})},\ \Eprint
    {http://arxiv.org/abs/1802.04847} {arXiv:1802.04847 [astro-ph.CO]}
    \BibitemShut {NoStop}%
    \bibitem [{\citenamefont {Bhatt}\ and\ \citenamefont
        {Pandey}(2016)}]{Pandey:2015kaa}%
    \BibitemOpen
    \bibfield  {author} {\bibinfo {author} {\bibfnamefont {J.~R.}\ \bibnamefont
            {Bhatt}}\ and\ \bibinfo {author} {\bibfnamefont {A.~K.}\ \bibnamefont
            {Pandey}},\ }\href {\doibase 10.1103/PhysRevD.94.043536} {\bibfield
        {journal} {\bibinfo  {journal} {Phys. Rev.}\ }\textbf {\bibinfo {volume}
            {D94}},\ \bibinfo {pages} {043536} (\bibinfo {year} {2016})},\ \Eprint
    {http://arxiv.org/abs/1503.01878} {arXiv:1503.01878 [astro-ph.CO]}
    \BibitemShut {NoStop}%
    \bibitem [{\citenamefont {Anand}\ \emph {et~al.}(2017)\citenamefont {Anand},
        \citenamefont {Bhatt},\ and\ \citenamefont {Pandey}}]{Anand:2017zpg}%
    \BibitemOpen
    \bibfield  {author} {\bibinfo {author} {\bibfnamefont {S.}~\bibnamefont
            {Anand}}, \bibinfo {author} {\bibfnamefont {J.~R.}\ \bibnamefont {Bhatt}}, \
        and\ \bibinfo {author} {\bibfnamefont {A.~K.}\ \bibnamefont {Pandey}},\
    }\href {\doibase 10.1088/1475-7516/2017/07/051} {\bibfield  {journal}
        {\bibinfo  {journal} {JCAP}\ }\textbf {\bibinfo {volume} {1707}},\ \bibinfo
        {pages} {051} (\bibinfo {year} {2017})},\ \Eprint
    {http://arxiv.org/abs/1705.03683} {arXiv:1705.03683 [astro-ph.CO]}
    \BibitemShut {NoStop}%
    \bibitem [{\citenamefont {Subramanian}(2016)}]{Subramanian:2015lua}%
    \BibitemOpen
    \bibfield  {author} {\bibinfo {author} {\bibfnamefont {K.}~\bibnamefont
            {Subramanian}},\ }\href {\doibase 10.1088/0034-4885/79/7/076901} {\bibfield
        {journal} {\bibinfo  {journal} {Rept. Prog. Phys.}\ }\textbf {\bibinfo
            {volume} {79}},\ \bibinfo {pages} {076901} (\bibinfo {year} {2016})},\
    \Eprint {http://arxiv.org/abs/1504.02311} {arXiv:1504.02311 [astro-ph.CO]}
    \BibitemShut {NoStop}%
    \bibitem [{\citenamefont {{Field}}(1958)}]{Field}%
    \BibitemOpen
    \bibfield  {author} {\bibinfo {author} {\bibfnamefont {G.~B.}\ \bibnamefont
            {{Field}}},\ }\href {\doibase 10.1109/JRPROC.1958.286741} {\bibfield
        {journal} {\bibinfo  {journal} {Proceedings of the IRE}\ }\textbf {\bibinfo
            {volume} {46}},\ \bibinfo {pages} {240} (\bibinfo {year} {1958})}\BibitemShut
    {NoStop}%
    \bibitem [{\citenamefont {Pritchard}\ and\ \citenamefont
        {Loeb}(2012)}]{Pritchard:2011xb}%
    \BibitemOpen
    \bibfield  {author} {\bibinfo {author} {\bibfnamefont {J.~R.}\ \bibnamefont
            {Pritchard}}\ and\ \bibinfo {author} {\bibfnamefont {A.}~\bibnamefont
            {Loeb}},\ }\href {\doibase 10.1088/0034-4885/75/8/086901} {\bibfield
        {journal} {\bibinfo  {journal} {Rept. Prog. Phys.}\ }\textbf {\bibinfo
            {volume} {75}},\ \bibinfo {pages} {086901} (\bibinfo {year} {2012})},\
    \Eprint {http://arxiv.org/abs/1109.6012} {arXiv:1109.6012 [astro-ph.CO]}
    \BibitemShut {NoStop}%
    \bibitem [{\citenamefont {Wouthuysen}(1952)}]{1952AJ.....57R..31W}%
    \BibitemOpen
    \bibfield  {author} {\bibinfo {author} {\bibfnamefont {S.~A.}\ \bibnamefont
            {Wouthuysen}},\ }\href {\doibase 10.1086/106661} {\bibfield  {journal}
        {\bibinfo  {journal} {apj}\ }\textbf {\bibinfo {volume} {57}},\ \bibinfo
        {pages} {31} (\bibinfo {year} {1952})}\BibitemShut {NoStop}%
    \bibitem [{\citenamefont {Field}(1959)}]{1959ApJ...129..536F}%
    \BibitemOpen
    \bibfield  {author} {\bibinfo {author} {\bibfnamefont {G.~B.}\ \bibnamefont
            {Field}},\ }\href {\doibase 10.1086/146653} {\bibfield  {journal} {\bibinfo
            {journal} {apj}\ }\textbf {\bibinfo {volume} {129}},\ \bibinfo {pages} {536}
        (\bibinfo {year} {1959})}\BibitemShut {NoStop}%
    \bibitem [{\citenamefont {Jedamzik}\ \emph {et~al.}(1998)\citenamefont
        {Jedamzik}, \citenamefont {Katalini\ifmmode~\acute{c}\else \'{c}\fi{}},\ and\
        \citenamefont {Olinto}}]{Jedamzik:1998kk}%
    \BibitemOpen
    \bibfield  {author} {\bibinfo {author} {\bibfnamefont {K.}~\bibnamefont
            {Jedamzik}}, \bibinfo {author} {\bibfnamefont {V.~c.~v.}\ \bibnamefont
            {Katalini\ifmmode~\acute{c}\else \'{c}\fi{}}}, \ and\ \bibinfo {author}
        {\bibfnamefont {A.~V.}\ \bibnamefont {Olinto}},\ }\href {\doibase
        10.1103/PhysRevD.57.3264} {\bibfield  {journal} {\bibinfo  {journal} {Phys.
                Rev. D}\ }\textbf {\bibinfo {volume} {57}},\ \bibinfo {pages} {3264}
        (\bibinfo {year} {1998})}\BibitemShut {NoStop}%
    \bibitem [{\citenamefont {Subramanian}\ and\ \citenamefont
        {Barrow}(1998)}]{Subramanian:1997gi}%
    \BibitemOpen
    \bibfield  {author} {\bibinfo {author} {\bibfnamefont {K.}~\bibnamefont
            {Subramanian}}\ and\ \bibinfo {author} {\bibfnamefont {J.~D.}\ \bibnamefont
            {Barrow}},\ }\href {\doibase 10.1103/PhysRevD.58.083502} {\bibfield
        {journal} {\bibinfo  {journal} {Phys. Rev.}\ }\textbf {\bibinfo {volume}
            {D58}},\ \bibinfo {pages} {083502} (\bibinfo {year} {1998})},\ \Eprint
    {http://arxiv.org/abs/astro-ph/9712083} {arXiv:astro-ph/9712083 [astro-ph]}
    \BibitemShut {NoStop}%
    \bibitem [{\citenamefont {Hogan}(1983)}]{Hogan:1983cj}%
    \BibitemOpen
    \bibfield  {author} {\bibinfo {author} {\bibfnamefont {C.~J.}\ \bibnamefont
            {Hogan}},\ }\href {\doibase 10.1103/PhysRevLett.51.1488} {\bibfield
        {journal} {\bibinfo  {journal} {Phys. Rev. Lett.}\ }\textbf {\bibinfo
            {volume} {51}},\ \bibinfo {pages} {1488} (\bibinfo {year}
        {1983})}\BibitemShut {NoStop}%
    \bibitem [{\citenamefont {Durrer}\ and\ \citenamefont
        {Caprini}(2003)}]{Durrer:2003ja}%
    \BibitemOpen
    \bibfield  {author} {\bibinfo {author} {\bibfnamefont {R.}~\bibnamefont
            {Durrer}}\ and\ \bibinfo {author} {\bibfnamefont {C.}~\bibnamefont
            {Caprini}},\ }\href {\doibase 10.1088/1475-7516/2003/11/010} {\bibfield
        {journal} {\bibinfo  {journal} {JCAP}\ }\textbf {\bibinfo {volume} {0311}},\
        \bibinfo {pages} {010} (\bibinfo {year} {2003})},\ \Eprint
    {http://arxiv.org/abs/astro-ph/0305059} {arXiv:astro-ph/0305059 [astro-ph]}
    \BibitemShut {NoStop}%
    \bibitem [{\citenamefont {Tashiro}\ and\ \citenamefont
        {Sugiyama}(2006{\natexlab{b}})}]{Tashiro:2006uv}%
    \BibitemOpen
    \bibfield  {author} {\bibinfo {author} {\bibfnamefont {H.}~\bibnamefont
            {Tashiro}}\ and\ \bibinfo {author} {\bibfnamefont {N.}~\bibnamefont
            {Sugiyama}},\ }\href {\doibase 10.1111/j.1365-2966.2006.10901.x} {\bibfield
        {journal} {\bibinfo  {journal} {Mon. Not. Roy. Astron. Soc.}\ }\textbf
        {\bibinfo {volume} {372}},\ \bibinfo {pages} {1060} (\bibinfo {year}
        {2006}{\natexlab{b}})},\ \Eprint {http://arxiv.org/abs/astro-ph/0607169}
    {arXiv:astro-ph/0607169 [astro-ph]} \BibitemShut {NoStop}%
    \bibitem [{\citenamefont {Aghanim}\ \emph {et~al.}(2018)\citenamefont {Aghanim}
        \emph {et~al.}}]{Aghanim:2018eyx}%
    \BibitemOpen
    \bibfield  {author} {\bibinfo {author} {\bibfnamefont {N.}~\bibnamefont
            {Aghanim}} \emph {et~al.} (\bibinfo {collaboration} {Planck}),\ }\href@noop
    {} {\  (\bibinfo {year} {2018})},\ \Eprint {http://arxiv.org/abs/1807.06209}
    {arXiv:1807.06209 [astro-ph.CO]} \BibitemShut {NoStop}%
    \bibitem [{\citenamefont {Cowling}(1956)}]{Cowling:1956gt}%
    \BibitemOpen
    \bibfield  {author} {\bibinfo {author} {\bibfnamefont {T.~G.}\ \bibnamefont
            {Cowling}},\ }\href {\doibase 10.1093/mnras/116.1.114} {\bibfield  {journal}
        {\bibinfo  {journal} {MNRAS}\ }\textbf {\bibinfo {volume} {116}},\ \bibinfo
        {pages} {114} (\bibinfo {year} {1956})}\BibitemShut {NoStop}%
    \bibitem [{\citenamefont {Shang}\ \emph {et~al.}(2002)\citenamefont {Shang},
        \citenamefont {Glassgold}, \citenamefont {Shu},\ and\ \citenamefont
        {Lizano}}]{Shang:2001df}%
    \BibitemOpen
    \bibfield  {author} {\bibinfo {author} {\bibfnamefont {H.}~\bibnamefont
            {Shang}}, \bibinfo {author} {\bibfnamefont {A.~E.}\ \bibnamefont
            {Glassgold}}, \bibinfo {author} {\bibfnamefont {F.~H.}\ \bibnamefont {Shu}},
        \ and\ \bibinfo {author} {\bibfnamefont {S.}~\bibnamefont {Lizano}},\ }\href
    {\doibase 10.1086/324197} {\bibfield  {journal} {\bibinfo  {journal}
            {Astrophys. J.}\ }\textbf {\bibinfo {volume} {564}},\ \bibinfo {pages} {853}
        (\bibinfo {year} {2002})},\ \Eprint {http://arxiv.org/abs/astro-ph/0110539}
    {arXiv:astro-ph/0110539 [astro-ph]} \BibitemShut {NoStop}%
    \bibitem [{\citenamefont {Draine}(1980)}]{Draine:1980bt}%
    \BibitemOpen
    \bibfield  {author} {\bibinfo {author} {\bibfnamefont {B.~T.}\ \bibnamefont
            {Draine}},\ }\href {\doibase 10.1086/158416} {\bibfield  {journal} {\bibinfo
            {journal} {APJ}\ }\textbf {\bibinfo {volume} {241}},\ \bibinfo {pages} {1021}
        (\bibinfo {year} {1980})}\BibitemShut {NoStop}%
    \bibitem [{\citenamefont {Schleicher}\ \emph {et~al.}(2009)\citenamefont
        {Schleicher}, \citenamefont {Banerjee},\ and\ \citenamefont
        {Klessen}}]{Schleicher:2008hc}%
    \BibitemOpen
    \bibfield  {author} {\bibinfo {author} {\bibfnamefont {D.~R.~G.}\
            \bibnamefont {Schleicher}}, \bibinfo {author} {\bibfnamefont
            {R.}~\bibnamefont {Banerjee}}, \ and\ \bibinfo {author} {\bibfnamefont
            {R.~S.}\ \bibnamefont {Klessen}},\ }\href {\doibase
        10.1088/0004-637X/692/1/236} {\bibfield  {journal} {\bibinfo  {journal}
            {Astrophys. J.}\ }\textbf {\bibinfo {volume} {692}},\ \bibinfo {pages} {236}
        (\bibinfo {year} {2009})},\ \Eprint {http://arxiv.org/abs/0808.1461}
    {arXiv:0808.1461 [astro-ph]} \BibitemShut {NoStop}%
    \bibitem [{\citenamefont {Mu\~noz}\ \emph {et~al.}(2015)\citenamefont
        {Mu\~noz}, \citenamefont {Kovetz},\ and\ \citenamefont
        {Ali-Ha\"{\i}moud}}]{Munoz:2015bk}%
    \BibitemOpen
    \bibfield  {author} {\bibinfo {author} {\bibfnamefont {J.~B.}\ \bibnamefont
            {Mu\~noz}}, \bibinfo {author} {\bibfnamefont {E.~D.}\ \bibnamefont {Kovetz}},
        \ and\ \bibinfo {author} {\bibfnamefont {Y.}~\bibnamefont
            {Ali-Ha\"{\i}moud}},\ }\href {\doibase 10.1103/PhysRevD.92.083528} {\bibfield
        {journal} {\bibinfo  {journal} {Phys. Rev. D}\ }\textbf {\bibinfo {volume}
            {92}},\ \bibinfo {pages} {083528} (\bibinfo {year} {2015})}\BibitemShut
    {NoStop}%
    \bibitem [{\citenamefont {Ali-Haimoud}\ and\ \citenamefont
        {Hirata}(2011)}]{AliHaimoud:2010dx}%
    \BibitemOpen
    \bibfield  {author} {\bibinfo {author} {\bibfnamefont {Y.}~\bibnamefont
            {Ali-Haimoud}}\ and\ \bibinfo {author} {\bibfnamefont {C.~M.}\ \bibnamefont
            {Hirata}},\ }\href {\doibase 10.1103/PhysRevD.83.043513} {\bibfield
        {journal} {\bibinfo  {journal} {Phys. Rev.}\ }\textbf {\bibinfo {volume}
            {D83}},\ \bibinfo {pages} {043513} (\bibinfo {year} {2011})},\ \Eprint
    {http://arxiv.org/abs/1011.3758} {arXiv:1011.3758 [astro-ph.CO]} \BibitemShut
    {NoStop}%
    \bibitem [{\citenamefont {Kovetz}\ \emph {et~al.}(2018)\citenamefont {Kovetz},
        \citenamefont {Poulin}, \citenamefont {Gluscevic}, \citenamefont {Boddy},
        \citenamefont {Barkana},\ and\ \citenamefont {Kamionkowski}}]{Kovetz:2018P}%
    \BibitemOpen
    \bibfield  {author} {\bibinfo {author} {\bibfnamefont {E.~D.}\ \bibnamefont
            {Kovetz}}, \bibinfo {author} {\bibfnamefont {V.}~\bibnamefont {Poulin}},
        \bibinfo {author} {\bibfnamefont {V.}~\bibnamefont {Gluscevic}}, \bibinfo
        {author} {\bibfnamefont {K.~K.}\ \bibnamefont {Boddy}}, \bibinfo {author}
        {\bibfnamefont {R.}~\bibnamefont {Barkana}}, \ and\ \bibinfo {author}
        {\bibfnamefont {M.}~\bibnamefont {Kamionkowski}},\ }\href {\doibase
        10.1103/PhysRevD.98.103529} {\bibfield  {journal} {\bibinfo  {journal} {Phys.
                Rev. D}\ }\textbf {\bibinfo {volume} {98}},\ \bibinfo {pages} {103529}
        (\bibinfo {year} {2018})}\BibitemShut {NoStop}%
    \bibitem [{\citenamefont {Boddy}\ \emph {et~al.}(2018)\citenamefont {Boddy},
        \citenamefont {Gluscevic}, \citenamefont {Poulin}, \citenamefont {Kovetz},
        \citenamefont {Kamionkowski},\ and\ \citenamefont {Barkana}}]{Boddy:2018G}%
    \BibitemOpen
    \bibfield  {author} {\bibinfo {author} {\bibfnamefont {K.~K.}\ \bibnamefont
            {Boddy}}, \bibinfo {author} {\bibfnamefont {V.}~\bibnamefont {Gluscevic}},
        \bibinfo {author} {\bibfnamefont {V.}~\bibnamefont {Poulin}}, \bibinfo
        {author} {\bibfnamefont {E.~D.}\ \bibnamefont {Kovetz}}, \bibinfo {author}
        {\bibfnamefont {M.}~\bibnamefont {Kamionkowski}}, \ and\ \bibinfo {author}
        {\bibfnamefont {R.}~\bibnamefont {Barkana}},\ }\href {\doibase
        10.1103/PhysRevD.98.123506} {\bibfield  {journal} {\bibinfo  {journal} {Phys.
                Rev. D}\ }\textbf {\bibinfo {volume} {98}},\ \bibinfo {pages} {123506}
        (\bibinfo {year} {2018})}\BibitemShut {NoStop}%
    \bibitem [{\citenamefont {Harker}\ \emph {et~al.}(2015)\citenamefont {Harker},
        \citenamefont {Mirocha}, \citenamefont {Burns},\ and\ \citenamefont
        {Pritchard}}]{Harker:2015M}%
    \BibitemOpen
    \bibfield  {author} {\bibinfo {author} {\bibfnamefont {G.~J.~A.}\
            \bibnamefont {Harker}}, \bibinfo {author} {\bibfnamefont {J.}~\bibnamefont
            {Mirocha}}, \bibinfo {author} {\bibfnamefont {J.~O.}\ \bibnamefont {Burns}},
        \ and\ \bibinfo {author} {\bibfnamefont {J.~R.}\ \bibnamefont {Pritchard}},\
    }\href {\doibase 10.1093/mnras/stv2630} {\bibfield  {journal} {\bibinfo
            {journal} {Monthly Notices of the Royal Astronomical Society}\ }\textbf
        {\bibinfo {volume} {455}},\ \bibinfo {pages} {3829} (\bibinfo {year}
        {2015})}\BibitemShut {NoStop}%
    \bibitem [{\citenamefont {Mirocha}\ \emph {et~al.}(2015)\citenamefont
        {Mirocha}, \citenamefont {Harker},\ and\ \citenamefont
        {Burns}}]{Mirocha:2015G}%
    \BibitemOpen
    \bibfield  {author} {\bibinfo {author} {\bibfnamefont {J.}~\bibnamefont
            {Mirocha}}, \bibinfo {author} {\bibfnamefont {G.~J.~A.}\ \bibnamefont
            {Harker}}, \ and\ \bibinfo {author} {\bibfnamefont {J.~O.}\ \bibnamefont
            {Burns}},\ }\href {\doibase 10.1088/0004-637x/813/1/11} {\bibfield  {journal}
        {\bibinfo  {journal} {The Astrophysical Journal}\ }\textbf {\bibinfo {volume}
            {813}},\ \bibinfo {pages} {11} (\bibinfo {year} {2015})}\BibitemShut
    {NoStop}%
    \bibitem [{\citenamefont {Zygelman}(2005)}]{Zygelman:2005}%
    \BibitemOpen
    \bibfield  {author} {\bibinfo {author} {\bibfnamefont {B.}~\bibnamefont
            {Zygelman}},\ }\href {\doibase 10.1086/427682} {\bibfield  {journal}
        {\bibinfo  {journal} {The Astrophysical Journal}\ }\textbf {\bibinfo {volume}
            {622}},\ \bibinfo {pages} {1356} (\bibinfo {year} {2005})}\BibitemShut
    {NoStop}%
\end{thebibliography}

\end{document}